\documentclass[draftclsnofoot,twoside,onecolumn,letter,12pt]{IEEEtran}
\usepackage{graphicx,cite,amssymb,amsmath,psfrag,subfigure}
\usepackage{graphicx}
\usepackage{amssymb}
\usepackage{amsmath}
\usepackage{cite}
\usepackage{subfigure}
\usepackage{mathrsfs}
\newtheorem{definition}{Definition}
\newtheorem{theorem}{Theorem}

\newtheorem{corollary}{Corollary}
\newtheorem{proposition}{Proposition}

\newtheorem{remark}{Remark}

\DeclareMathOperator{\E}{\mathbb{E}}

\newcommand{\EXs}[2]{\E_{{#1}}\left\{{#2}\right\}}
\newcommand{\PDF}[2]{p_{{#1}}\left({#2}\right)}
\newcommand{\CG}[2]{\mathcal{CN}\left({#1},{#2}\right)}
\newcommand{\Erv}[1]{\mathcal{E}\left({#1}\right)}

\newcommand{\PR}{p_{\mathrm{R}}}
\newcommand{\B}[1]{{\pmb{#1}}}

\newcommand{\T}{\mathcal{T}}

\newcommand{\Link}[1]{{\mathrm{L}_{#1}}}
\newcommand{\Pe}[1]{P_\mathrm{e}^{\left(#1\right)}}
\newcommand{\Ro}[1]{R_{0,#1}}
\newcommand{\mPe}{P_\mathrm{e}^\star}
\newcommand{\Rcr}[1]{R_{\mathrm{cr},#1}}
\newcommand{\SR}{\mathsf{R}}

\newcommand{\FoxH}[7]{H^{#1,#2}_{#3,#4}
                        \left[
                            {#5}
                            \left|
                            \begin{array}{l}
                                {#6}
                                \\
                                {#7}
                            \end{array}
                            \right.
                        \right]}

\newcommand{\GFoxH}[8]{H^{#1}_{#2}
                        \left[
                            \begin{matrix}
                            {#3} \\[3mm] {#4}
                            \end{matrix}
                            \left|
                            \begin{array}{l}
                                {#5}
                                \\
                                {#6}
                                \\
                                {#7}
                                \\
                                {#8}
                            \end{array}
                            \right.
                        \right]}

\makeatletter
\def\@setsize#1#2#3#4{
    \@nomath#1
    \let\@currsize#1
    \baselineskip #2
    \baselineskip \baselinestretch\baselineskip
    \parskip \baselinestretch\parskip
    \setbox\strutbox \hbox{
        \vrule height.7\baselineskip
            depth.3\baselineskip
            width\z@}
    \skip\footins \baselinestretch\skip\footins
    \normalbaselineskip\baselineskip#3#4}
\makeatother

\makeatletter
\newcommand{\setstretch}[1]{
    \def\baselinestretch{#1}%
    \@currsize
    }
\makeatother

\makeatletter
\newenvironment{salign}{
    \vskip -0.5\baselineskip   
    \setstretch{1}
    \start@align\@ne\st@rredfalse\m@ne
    }{
    \math@cr
    \black@\totwidth@
    \egroup
    \ifingather@
        \restorealignstate@
        \egroup
        \nonumber
        \ifnum0=`{\fi\iffalse}\fi
        \else
        $$%
    \fi
    \ignorespacesafterend
    \vspace{-6mm}  
    \vskip 0.6\baselineskip
    \par
    \noindent
    }
\makeatother

\def\BibTeX{{\rm B\kern-.05em{\sc i\kern-.025em b}\kern-.08em
    T\kern-.1667em\lower.7ex\hbox{E}\kern-.125emX}}

\begin{document}

\title{
    \hspace{4cm}\\[-0.7cm]
    Two-Way Relay Channels: Error Exponents and Resource
    Allocation
}
\author{
        Hien Quoc Ngo, \IEEEmembership{Student Member, IEEE},
        Tony Q. S. Quek, \IEEEmembership{Member, IEEE},
        and
        Hyundong Shin, \IEEEmembership{Member, IEEE}
\thanks{
        H.~Q.\ Ngo and H.\ Shin are with the Department of Electronics and Radio Engineering, Kyung Hee University, 1 Seocheon-dong, Giheung-gu, Yongin-si,
        Gyeonggi-do 446-701, Korea
        (Email: ngoquochien@khu.ac.kr; hshin@khu.ac.kr).
}
\thanks{
        T.~Q.~S.\ Quek is with the Institute for Infocomm Research, 1 Fusionopolis Way, \#21-01 Connexis South Tower, Singapore 138632
        (Email: qsquek@ieee.org).
}
}

\markboth{Submitted to the IEEE Transactions on Communications}
        {Ngo \textit{\MakeLowercase{et al.}}:
        Two-Way Relay Channels: Error Exponents and Resource
        Allocation}

\maketitle

\renewcommand{\baselinestretch}{1.5} \normalsize
\vspace{-1.0cm}

\begin{abstract}

In a two-way relay network, two terminals exchange information
over a shared wireless half-duplex channel with the help of a
relay. Due to its fundamental and practical importance, there has
been an increasing interest in this channel. However,
surprisingly, there has been little work that characterizes the
fundamental tradeoff between the communication reliability and
transmission rate across all signal-to-noise ratios. In this
paper, we consider amplify-and-forward (AF) two-way relaying due
to its simplicity. We first derive the random coding error
exponent for the link in each direction. From the exponent
expression, the capacity and cutoff rate for each link are also
deduced. We then put forth the notion of the \emph{bottleneck}
error exponent, which is the worst exponent decay between the two
links, to give us insight into the fundamental tradeoff between
the rate pair and information-exchange reliability of the two
terminals. As applications of the error exponent analysis, we
present two optimal resource allocations to maximize the
bottleneck error exponent: i) the optimal rate allocation under a
sum-rate constraint and its closed-form \emph{quasi-optimal}
solution that requires only knowledge of the capacity and cutoff
rate of each link; and ii) the optimal power allocation under a
total power constraint, which is formulated as a quasi-convex
optimization problem. Numerical results verify our analysis and
the effectiveness of the optimal rate and power allocations in
maximizing the bottleneck error exponent.

\end{abstract}


\clearpage

\section{Introduction}

\begin{figure}[t]
    \centerline{\includegraphics[width=0.85\textwidth]{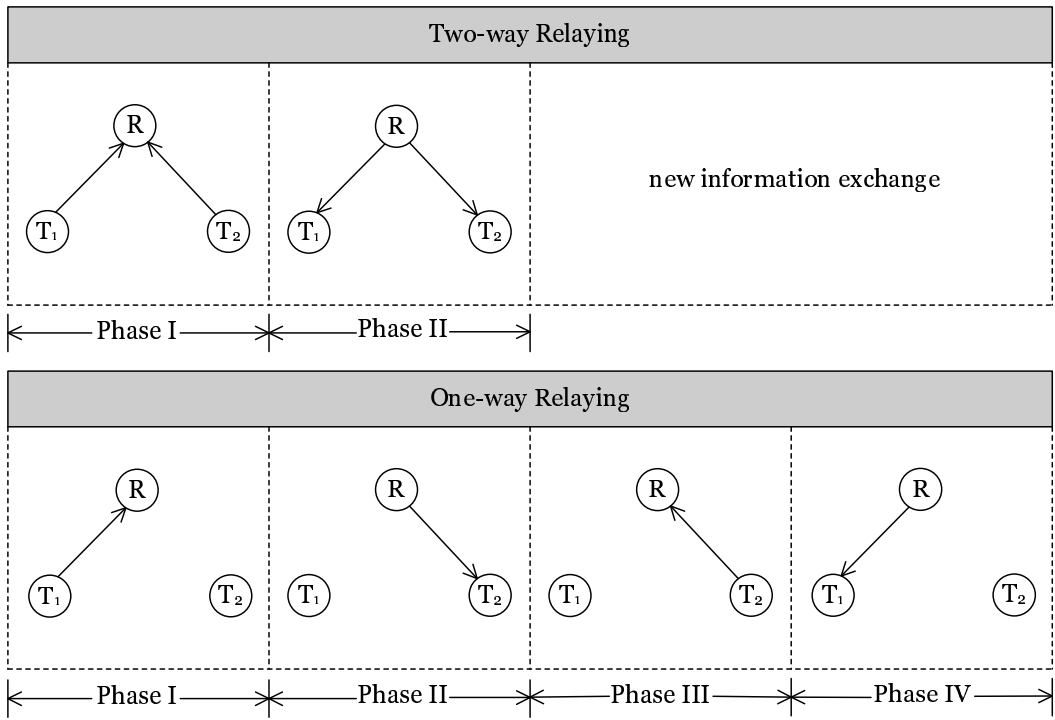}}
    \caption{Information exchange with one- and two-way relaying.}
    \label{fig: model}
\end{figure}

The two-way communication channel was first introduced by Shannon,
showing how to efficiently design message structures to enable
simultaneous bidirectional communication at the highest possible
data rates \cite{Sha:61:BSPS}. Recently, this model has regained
significant interest by introducing an additional relay to support
the exchange of information between the two communicating
terminals. The attractive feature of this two-way relay model is
that it can compensate the spectral inefficiency of one-way
relaying under a half-duplex constraint
\cite{RW:07:JSAC,PY:07:ICC,KMT:08:IT,OBSB:08:IT,SSO:08:ITW,CHK:08:ICC}.
With one-way relaying, we should use four phases to exchange
information between two terminals via a half-duplex relay, i.e.,
it takes two phases to send information from one terminal to the
other terminal and two phases for the reverse direction (see
Fig.~\ref{fig: model}). However, exploiting the knowledge of
terminals' own transmitted signals and the broadcast nature of the
wireless medium, we can improve the spectral efficiency by using
only two phases to exchange information in the two-way relay
channel (TWRC) \cite{RW:07:JSAC}.

Due to the aforementioned fundamental and practical importance of
the TWRC, much work has investigated the sum rate and the
achievable rate region of the TWRC with different relaying
protocols
\cite{RW:07:JSAC,PY:07:ICC,KMT:08:IT,OBSB:08:IT,SSO:08:ITW,CHK:08:ICC}.
The half-duplex amplify-and-forward (AF) and decode-and-forward
(DF) TWRCs have been studied in \cite{RW:07:JSAC} where it was
shown that both protocols with two-way relaying can redeem a
significant portion of the half-duplex loss. In \cite{PY:07:ICC},
the achievable rates for AF, DF, joint-DF, and denoise-and-forward
relaying have been analyzed and the condition for maximization of
the two-way rate are investigated for each relaying scheme. The
broadcast capacity region in terms of the maximal probability of
error has been derived in \cite{OBSB:08:IT} for the DF TWRC. A new
achievable rate region for the TWRC has been found in
\cite{SSO:08:ITW} for partial DF relaying, which is a
superposition of both DF and compress-and-forward relaying. Bit
error probability at each terminal has also been analyzed for a
memoryless additive white Gaussian noise (AWGN) TWRC
\cite{CHK:08:ICC}. However, there has been few work that
characterizes the fundamental tradeoff between the communication
reliability and transmission rate in the TWRC across all
signal-to-noise ratio (SNR) regimes.

In this paper, we consider half-duplex AF two-way relaying due to
its simplicity in practical implementation. To characterize the
fundamental tradeoff between the communication reliability and
rate, we first derive Gallager's random coding error exponent
(RCEE)---the classical lower bound to Shannon's reliability
function (see, e.g.,
\cite{Gal:68:Book,Ahm:97:PhD,AM:99:IT,SW:05:COM} and references
therein)---for the link of each direction in the AF
TWRC.\footnote{
        In the following, we shall use simply the term ``TWRC'' to
        denote the AF TWRC.
        }
Instead of considering only the achievable rate or error
probability as a performance measure, the RCEE results can reveal
the inherent tradeoff between these measures to unveil the
effectiveness of two-way relaying in redeeming a significant
portion of the half-duplex loss in the information exchange. From
the exponent expression, the capacity and cutoff rate for each
link in the TWRC are further deduced. We then introduce the
\emph{bottleneck} error exponent, which is defined by the worst
exponent decay between the links of two directions, to capture the
tradeoff between the rate pair of both links and the reliability
of information exchange at such a rate pair. Using this notion, we
can appertain a bottleneck exponent value to each rate pair and
characterize the bottleneck exponent plane from the set of all
possible rate pairs besides the achievable rate region. This
enables us to design a two-way relay network with reliable
information exchange.

For applications of the error exponent analysis for the TWRC, we
present two optimal resource (rate and power) allocations, the
main results of which can be summarized as follows.

\begin{itemize}

\item

We show that the optimal rate allocation to maximize the
bottleneck error exponent under a sum-rate constraint is a rate
pair such that the RCEE values of both links become identical at
the respective rates. This optimal rate pair can be determined by
a closed-form solution for sum rates less than a certain
constant---called the \emph{decisive sum rate}---depending only on
the cutoff and critical rates of each link. Furthermore, the
optimal solution requires only the knowledge of each cutoff rate.
At sum rates larger than the decisive point, we can allocate a
rate pair \emph{quasi-optimally} in closed form, requiring only
knowledge of the capacity and cutoff rate of each link.

\item

We determine the optimal power allocation that maximizes the
bottleneck error exponent under a total power constraint of the
two terminals. In the presence of perfect global channel state
information (CSI), we show that this power allocation problem can
be formulated as a quasi-convex optimization problem, where the
optimal solution can be efficiently determined via a sequence of
convex feasibility problems in the form of second-order cone
programs (SOCPs).

\end{itemize}

The rest of this paper is organized as follows. In Section II, we
describe the system model. In Section III, we present the results
of the error exponent analysis for the TWRC. The optimization
framework for two-way relay networks is developed for the rate and
power allocations to maximize the bottleneck error exponent in
Section IV. We provide some numerical results in Section V and
finally conclude the paper in Section VI.

\textit{Notation:} Throughout the paper, we shall use the
following notation. Boldface upper- and lower-case letters denote
matrices and column vectors, respectively. The superscript
$(\cdot)^{T}$ denotes the transpose. We use $\mathbb{R}$,
$\mathbb{R}_{+}$, and $\mathbb{R}_{++}$ to denote the set of real
numbers, nonnegative real numbers, and positive real numbers,
respectively. A circularly symmetric complex Gaussian distribution
with mean $\mu$ and variance $\sigma^2$ is denoted by
$\CG{\mu}{\sigma^2}$ and the exponential distribution with a
hazard rate $\lambda$ is denoted by $\Erv{\lambda}$.

\section{System Model}

We consider the TWRC as illustrated in Fig.~\ref{fig: model},
where a half-duplex relay node $\mathrm{R}$ bidirectionally
communicates between two terminals $\mathrm{T}_{k \in
\T=\left\{1,2\right\}}$ with AF relaying. In the first multiple
access phase, the terminals $\mathrm{T}_{k \in \T}$ transmit their
information to the relay and the received signal at
the relay is given by 
%
%
\begin{align}\label{eq:sys_two_1}
    y_{\mathrm{R}}
    &=
        h_{1} x_{1} + h_{2} x_{2} + z_{\mathrm{R}}
\end{align}
where $x_{k \in \T}$ is the transmitted signal from the terminal
$\mathrm{T}_{k \in \T}$ with
$\mathbb{E}\bigl\{|x_{k}|^2\bigr\}=p_k$ , $h_{k} \sim
\CG{0}{\Omega_{k}}$ is the channel coefficient from
$\mathrm{T}_{k}$ to the relay, and $z_{\mathrm{R}} \sim
\CG{0}{N_{0}}$ is the complex AWGN.\footnote{
        We assume the channel reciprocity for $h_k$ as in \cite{RW:07:JSAC}.
        }
Note that $|h_{k}|^2 \sim \Erv{1/\Omega_{k}}$.

At the relay, the received signal is scaled and broadcasted to
both terminals in the second broadcast phase, while satisfying its
power constraint $p_\mathrm{R}$. Then, the received signal at the
terminal $\mathrm{T}_{k \in \T}$ is given by
\begin{align}\label{eq:sys_two_2}
    y_{\mathrm{T},k}
    &=
        h_{k} x_{\mathrm{R}} + z_{\mathrm{T},k}
\end{align}
where $x_{\mathrm{R}} = G y_{\mathrm{R}} $ is the transmitted
signal from the relay with
$\mathbb{E}\bigl\{|x_\mathrm{R}|^2\bigr\}=p_\mathrm{R}$,
$z_{\mathrm{T},k} \sim \CG{0}{N_{0}}$ is the AWGN, and $G$ is the
relaying gain given by
\begin{align}\label{eq:sys_two_21}
    G
    =
        \sqrt{\frac{\PR}{p_{1}|h_{1}|^2 + p_{2}|h_{2}|^{2} + N_{0}}}.
\end{align}
We impose a total transmit power constraint $P$ such that $p_1 +
p_2 \leq P$. As in \cite{RW:07:JSAC}, we further assume that the
terminal $\mathrm{T}_{k \in \T}$ knows its own transmitted signal
and has perfect CSI to remove self-interference prior to decoding.
For notational convenience, we shall refer to the communication
link $\mathrm{T}_1 \rightarrow \mathrm{R} \rightarrow
\mathrm{T}_2$ as the link $\Link{1}$ and $\mathrm{T}_2 \rightarrow
\mathrm{R} \rightarrow \mathrm{T}_1$ as the link $\Link{2}$,
respectively. With the self-interference cancellation, the
effective SNR of the link $\Link{k \in \T}$ is given by
\begin{align}
        \label{eq:sys_two_4}
    \gamma_{k}^{\mathrm{eff}}
    &=
        \frac{p_{k} \PR \alpha_{1} \alpha_{2}}{p_{k}\alpha_{k} + \left(\PR +p_{1}p_{2}/p_{k}\right)\alpha_{1}\alpha_{2}/\alpha_{k} + 1}
\end{align}
where $\alpha_{k\in \T} \triangleq |h_{k}|^{2}/N_{0}$.

\section{Error Exponent Analysis}

\subsection{Mathematical Preliminaries}

The reliability function or error exponent for a channel of the
capacity $C$ is the best exponent decay with the codeword length
$N$ in the average probability of error that one can achieve at a
rate $R<C$ \cite{Gal:68:Book,Ahm:97:PhD,AM:99:IT,SW:05:COM}:
\begin{align}\label{eq:prelim_1}
    E\left(R\right)
    &\triangleq
        \limsup_{N \rightarrow \infty}
            -\frac{1}{N}
            \ln P_\mathrm{e}^{\mathrm{opt}}\left(R,N\right)
\end{align}
where $P_\mathrm{e}^{\mathrm{opt}}\left(R,N\right)$ is the average
block error probability for the optimal block code of length $N$
and rate $R$.\footnote{
     Throughout the paper, we shall use a rate measured in units of nats per second per Hz (nats/s/Hz).
     }
As a classical lower bound on the reliability function, the RCEE
or Gallager's exponent is given by \cite{Gal:68:Book}
\begin{align}\label{eq:prelim_5}
    E_{\mathrm{r}}\left(R\right)
    &\triangleq
        \max_{Q}
        \max_{0 \leq \rho \leq 1}
        \left\{
            E_{0}\right(\rho,Q\left)
            - \rho R
        \right\}
\end{align}
with
\begin{align}  \label{eq:prelim_4}
    E_{0}\left(\rho,Q\right)
    &\triangleq
        - \ln\left\{
            \int
            \left[
                \int
                    Q\left(x\right)
                    p\left(y|x\right)^{\frac{1}{1 + \rho}}
                    dx
            \right]^{1 + \rho}
            dy
            \right\}
\end{align}
where $Q\left(x\right)$ is the input distribution and
$p\left(y|x\right)$ is the transition probability. Unfortunately,
the double maximization in \eqref{eq:prelim_5} is generally very
difficult since the inner integral is raised to a fractional
exponent when $\rho \in \left(0,1\right)$ and the lack of
knowledge about the optimal input distribution $Q\left(x\right)$.
For analytical tractability, the Gaussian input distribution
$Q(x)$ is often used, which is optimal if the rate $R$ approaches
the channel capacity
\cite{Gal:68:Book,Ahm:97:PhD,AM:99:IT,SW:05:COM}.

\subsection{Two-way Relay Channels}

The TWRC consists of two communication links and the achievable
rate can be characterized by the sum rate of two parallel relay
channels under perfect self-interference cancellation
\cite{RW:07:JSAC}. As such, we need to first consider the RCEE for
each link and subsequently introduce a notion of the bottleneck
error exponent for the TWRC to effectively capture the tradeoff
between the individual rates and the reliability. Using the
Gaussian input distribution, we obtain the following proposition
for the RCEE of each link in the TWRC.


\begin{proposition}\label{prop1}

With the Gaussian input distribution, the RCEE for the link
$\Link{k \in \T}$ of the TWRC with AF relaying is given by
\begin{align}\label{eq:prop1_1}
    E_{\mathrm{r},k}\left(R\right)
    &=
        \max_{\rho \in \left[0,1\right]}
            \left\{
                E_{\mathrm{0},k}\left(\rho\right)
                - 2\rho R
            \right\}
\end{align}
where
\begin{align}   \label{eq:prop1_2}
    E_{0,k}\left(\rho\right)
    &=
        - \ln
          \mathbb{E}_{\gamma^{\mathrm{eff}}_{k}}
          \left\{
            \left(
                1 + \frac{\gamma^{\mathrm{eff}}_{k}}{1 + \rho}
            \right)^{-\rho}
          \right\}.
\end{align}

\begin{proof}
It follows immediately from the results of \cite{AM:99:IT} along
with the self-interference cancellation at the terminal
$\mathrm{T}_{k \in \T}$.
\end{proof}

\end{proposition}


\begin{remark}

It is difficult to obtain a closed-form solution for
\eqref{eq:prop1_2} in Proposition~\ref{prop1} due to an
analytically intractable form of the effective SNR
$\gamma^{\mathrm{eff}}_{k \in \T}$  given in \eqref{eq:sys_two_4}.
In what follows, to alleviate such difficulty and render
\eqref{eq:prop1_2} more amenable to further analysis, we use the
upper bound $\gamma_k^{\mathrm{ub}}$ on the effective SNR
$\gamma^{\mathrm{eff}}_{k}$ by ignoring the term $1$ in the
denominator:
\begin{align}
    \gamma_{k}^{\mathrm{ub}}
    &=
        \frac{p_{k} \PR \alpha_{1} \alpha_{2}}{p_{k}\alpha_{k} + \left(\PR +p_1 p_{2}/p_k\right)\alpha_1\alpha_{2}/\alpha_k } \label{eq:RCE_1}
\end{align}
which corresponds to the ideal/hypothetical AF relaying
\cite{HA:03:WCOM,HA:04:COM,AK:04:WCOM}.

\end{remark}

\begin{remark}  \label{remark:2}

The factor $2$ of $\rho R$ in \eqref{eq:prop1_1} is due to the use
of two phases for the exchange of information in the TWRC. In
contrast, with one-way relaying, the information exchange occurs
over four phases and hence, this factor should be $4$, leading the
RCEE for each link to $E_{\mathrm{r},k}\left(R\right)= \max_{\rho
\in \left[0,1\right]} \left\{E_{\mathrm{0},k}\left(\rho\right) -
4\rho R \right\}$.\footnote{
        In the one-way relay channel (OWRC), if the total relaying power
        for information exchange is again constrained to $p_\mathrm{R}$,
        then the ideal/hypothetical AF relaying yields the upper bound on
        the effective SNR for the link
        $\Link{k \in \T}$ as 
        %
        \begin{align*}
            \gamma_{k}^{\mathrm{up}}
            =
                \frac{p_{k} \left(\PR/2\right) \alpha_{1}\alpha_{2}}{p_{k}\alpha_{k} + \left(\PR/2\right)\alpha_{1}\alpha_{2}/\alpha_{k}}
        \end{align*}
        which is slightly different from \eqref{eq:RCE_1} but makes no deviation in the analysis.
        }

\end{remark}


\begin{theorem}\label{thm1}

With the Gaussian input distribution, the RCEE for the link
$\Link{k \in \T}$ of the TWRC with ideal/hypothetical AF relaying
is given by
\begin{align} \label{eq:thm1 1}
        \tilde{E}_{\mathrm{r},k}\left(R\right)
    =
        \max_{\rho \in \left[0,1\right]}
            \left\{
                \tilde{E}_{0,k}\left(\rho\right)
                - 2\rho R
            \right\}
\end{align}
with
\begin{salign} \label{eq:thm1 3}
    \tilde{E}_{0,k}\left(\rho\right)
    &=
        -\ln
        \mathbb{E}_{\gamma^{\mathrm{ub}}_{k}}
        \left\{
            \left(
                1 + \frac{\gamma^{\mathrm{ub}}_{k}}{1 + \rho}
            \right)^{-\rho}
        \right\}
    \nonumber \\
    &=
        -\ln
        \left\{
            \frac{
                4\lambda_{k}
                \mu_{k}
                }{
                \sqrt{\pi}
                \Gamma\left(\rho\right)
                \left(
                    \sqrt{\lambda_{k}}
                    +\sqrt{\mu_{k}}
                \right)^{4}
                }
                \GFoxH{1,1,1,1,2}
                    {1,\left(1:1\right),0,\left(1:2\right)}
                    {\frac{\eta_{k}}{\left(1+\rho\right)\left(\sqrt{\lambda_{k}}+\sqrt{\mu_{k}}\right)^{2}}}
                    {\frac{4\sqrt{\lambda_{k} \mu_{k}}}{\left(\sqrt{\lambda_{k}}+\sqrt{\mu_{k}}\right)^{2}}}
                    {\left(2,1\right)}
                    {\left(1-\rho,1\right);\left(1/2,1\right)}
                    {\hspace{1.5cm}\text{---}}
                    {\left(0,1\right);\left(0,1\right),\left(0,1\right)}
        \right.
    \nonumber\\
    &\hspace{0.5cm}
        -
        \left.
            \frac{
                2\left(
                    \lambda_{k}+\mu_{k}\right)
                    \sqrt{\lambda_{k} \mu_{k}}
                }{
                \sqrt{\pi}
                \Gamma\left(\rho\right)
                \left(
                    \sqrt{\lambda_{k}}
                    +\sqrt{\mu_{k}}
                \right)^{4}
                }
                \GFoxH{1,1,1,1,2}
                    {1,\left(1:1\right),0,\left(1:2\right)}
                    {\frac{\eta_{k}}{\left(1+\rho\right)\left(\sqrt{\lambda_{k}}+\sqrt{\mu_{k}}\right)^{2}}}
                    {\frac{4\sqrt{\lambda_{k} \mu_{k}}}{\left(\sqrt{\lambda_{k}}+\sqrt{\mu_{k}}\right)^{2}}}
                    {\left(2,1\right)}
                    {\left(1-\rho,1\right);\left(1/2,1\right)}
                    {\hspace{1.5cm}\text{---}}
                    {\left(0,1\right);\left(1,1\right),\left(-1,1\right)}
        \right\}
    \nonumber \\
    &\hspace{11.5cm}
    \text{for $0 < \rho \leq 1$}
\end{salign}
and $\tilde{E}_{0,k}\left(\rho\right)=0$ for $\rho=0$, where
$\Gamma\left(\cdot\right)$ is Euler's gamma function,
$H^{K,N,N',M,M'}_{E,\left(A:C\right),F,\left(B:D\right)}\left[\cdot\right]$
is the generalized Fox $H$-function \cite[eq (2.2.1)]{MS:78:Book},
and
\begin{align*}
    \eta_{k}
    &=
        \frac{p_\mathrm{R}}{p_\mathrm{R}+p_{1}p_2/p_k}
    \\
    \lambda_{k}
    &=
        \frac{N_{0}}{p_{k} \Omega_{k}}
    \\
    \mu_{k}
    &=
        \frac{N_{0}\Omega_{k}}{\left(p_{1}p_{2}/p_{k}+ p_\mathrm{R}\right)
        \Omega_{1}\Omega_{2}}.
\end{align*}

\begin{proof}
See Appendix~\ref{sec:proof:thm1}.
\end{proof}

\end{theorem}


\begin{remark}

The maximum of the exponent
$\tilde{E}_{\mathrm{r},k}\left(R\right)$ over $\rho$ occurs at
$R=\frac{1}{2} \left.\left[ \partial
\tilde{E}_{0,k}\left(\rho\right) /\partial
\rho\right]\right|_{\rho=\rho_\mathrm{opt}}$ and hence, the slope
of the exponent--rate curve at a rate $R$ is equal to
$-2\rho_\mathrm{opt}$. The maximizing $\rho_\mathrm{opt}$ lies in
$\left[0,1\right]$ if
\begin{align}
    \Rcr{k}
    =
        \frac{1}{2} \!
        \left.
        \left[
            \frac{\partial \tilde{E}_{0,k}\left(\rho\right)}{\partial \rho}
        \right]
        \right|_{\rho=1}
    \leq
        R
    \leq
        \frac{1}{2}
        \left.
        \left[
            \frac{\partial \tilde{E}_{0,k}\left(\rho\right)}{\partial \rho}
        \right]
        \right|_{\rho=0}
    =
        \langle C_k \rangle
\end{align}
where $\Rcr{k}$ and $\langle C_k \rangle$ are the critical rate
and the (ergodic) capacity for the link $\Link{k \in \T}$,
respectively. For $R < \Rcr{k}$, we have $\rho_\mathrm{opt}=1$,
yielding the slope of the exponent--rate curve is equal to $-2$
and
$\tilde{E}_{\mathrm{r},k}\left(R\right)=\tilde{E}_{0,k}\left(1\right)-2R$.
Furthermore, the cutoff rate for the link $\Link{k \in \T}$ is
given by $\Ro{k}=\tilde{E}_{0,k}\left(1\right)/2$. This quantity
is equal to the value of $R$ at which the exponent becomes zero by
setting $\rho=1$. While the capacity determines the maximum
achievable rate, the cutoff rate determines the maximum practical
transmission rate for possible sequential decoding strategies and
indicates both the values of the zero-rate exponent and the rate
regime in which the error probability can be made arbitrarily
small by increasing the codeword length.

\end{remark}


\begin {corollary} \label {coro1: capacity}

The ergodic capacity for the link $\Link{k \in \T}$ of the TWRC
with ideal/hypothetical AF relaying is given by
\begin{salign} \label{eq:coro1 1}
    \langle C_{k} \rangle
    &=
            \frac{
                2\lambda_{k}
                \mu_{k}
                }{
                \sqrt{\pi}
                \left(
                    \sqrt{\lambda_{k}}
                    +\sqrt{\mu_{k}}
                \right)^{4}
                }
                \GFoxH{1,2,1,1,2}
                    {1,\left(2:1\right),0,\left(2:2\right)}
                    {\frac{\eta_{k}}{\left(\sqrt{\lambda_{k}}+\sqrt{\mu_{k}}\right)^{2}}}
                    {\frac{4\sqrt{\lambda_{k} \mu_{k}}}{\left(\sqrt{\lambda_{k}}+\sqrt{\mu_{k}}\right)^{2}}}
                    {\left(2,1\right)}
                    {\left(1,1\right),\left(1,1\right);\left(1/2,1\right)}
                    {\hspace{2cm}\text{---}}
                    {\left(1,1\right),\left(0,1\right);\left(0,1\right),\left(0,1\right)}
    \nonumber\\
    &\hspace{0.2cm}
        -
            \frac{
                \left(
                    \lambda_{k}+\mu_{k}\right)
                    \sqrt{\lambda_{k} \mu_{k}}
                }{
                \sqrt{\pi}
                \left(
                    \sqrt{\lambda_{k}}
                    +\sqrt{\mu_{k}}
                \right)^{4}
                }
                \GFoxH{1,2,1,1,2}
                    {1,\left(2:1\right),0,\left(2:2\right)}
                    {\frac{\eta_{k}}{\left(\sqrt{\lambda_{k}}+\sqrt{\mu_{k}}\right)^{2}}}
                    {\frac{4\sqrt{\lambda_{k} \mu_{k}}}{\left(\sqrt{\lambda_{k}}+\sqrt{\mu_{k}}\right)^{2}}}
                    {\left(2,1\right)}
                    {\left(1,1\right),\left(1,1\right);\left(1/2,1\right)}
                    {\hspace{2cm}\text{---}}
                    {\left(1,1\right),\left(0,1\right);\left(1,1\right),\left(-1,1\right)}.
\end{salign}

\begin{proof}
See Appendix~\ref{sec:proof:cor1}.
\end{proof}

\end {corollary}

\begin {corollary} \label {coro2: cutoff rate}

The cutoff rate for the link $\Link{k \in \T}$ of the TWRC with
ideal/hypothetical AF relaying is given by
\begin{salign}\label{eq:coro2 1}
    \Ro{k}
    &=
        -\frac{1}{2}\ln
        \left\{
            \frac{
                4\lambda_{k}
                \mu_{k}
                }{
                \sqrt{\pi}
                \left(
                    \sqrt{\lambda_{k}}
                    +\sqrt{\mu_{k}}
                \right)^{4}
                }
                \GFoxH{1,1,1,1,2}
                    {1,\left(1:1\right),0,\left(1:2\right)}
                    {\frac{\eta_{k}}{2\left(\sqrt{\lambda_{k}}+\sqrt{\mu_{k}}\right)^{2}}}
                    {\frac{4\sqrt{\lambda_{k} \mu_{k}}}{\left(\sqrt{\lambda_{k}}+\sqrt{\mu_{k}}\right)^{2}}}
                    {\left(2,1\right)}
                    {\left(0,1\right);\left(1/2,1\right)}
                    {\hspace{1.5cm}\text{---}}
                    {\left(0,1\right);\left(0,1\right),\left(0,1\right)}
        \right.
    \nonumber\\
    &\hspace{1.5cm}
        -
        \left.
            \frac{
                2\left(
                    \lambda_{k}+\mu_{k}\right)
                    \sqrt{\lambda_{k} \mu_{k}}
                }{
                \sqrt{\pi}
                \left(
                    \sqrt{\lambda_{k}}
                    +\sqrt{\mu_{k}}
                \right)^{4}
                }
                \GFoxH{1,1,1,1,2}
                    {1,\left(1:1\right),0,\left(1:2\right)}
                    {\frac{\eta_{k}}{2\left(\sqrt{\lambda_{k}}+\sqrt{\mu_{k}}\right)^{2}}}
                    {\frac{4\sqrt{\lambda_{k} \mu_{k}}}{\left(\sqrt{\lambda_{k}}+\sqrt{\mu_{k}}\right)^{2}}}
                    {\left(2,1\right)}
                    {\left(0,1\right);\left(1/2,1\right)}
                    {\hspace{1.5cm}\text{---}}
                    {\left(0,1\right);\left(1,1\right),\left(-1,1\right)}
        \right\}.
\end{salign}

\begin{proof}
It follows immediately from \eqref{eq:thm1 3} by setting $\rho=1$.
\end{proof}

\end{corollary}

\begin{remark}

It is insufficient to characterize the information \emph{exchange}
in the TWRC by only investigating the RCEE for each link
individually, as it just reflects the tradeoff between the
communication rate and reliability for the information
\emph{transmission} in one direction. Therefore, we introduce a
notion of the \emph{bottleneck exponent} for the TWRC to capture
the tradeoff between the rate pair of both links and the
reliability of information exchange at such a rate pair, enabling
us to optimize the resource allocation in the TWRC.

\end{remark}


\begin{definition} [Bottleneck Error Probability] \label{def: AvPe}

For a TWRC with the terminal $\mathrm{T}_{k \in \T}$ transmitting
a code $\left(e^{N R_k},N \right)$ of rate $R_k$, the bottleneck
error probability is defined as
\begin{align} \label{eq:def 1}
    \mPe
    \triangleq
        \max_{k \in \T} \Pe{k}
\end{align}
where $\Pe{k}$ is the error probability of the link $\Link{k}$.
\end{definition}

Note that Definition~\ref{def: AvPe} can be applicable for a
general TWRC, regardless of relaying protocols. From the random
coding bound
\begin{align} \label{eq:Pe 1}
    \Pe{k}
    \leq
        e^{-N \tilde{E}_{\mathrm{r},k}\left(R_{k}\right)}
\end{align}
the bottleneck error probability of the TWRC is bounded by
\begin{align} \label{eq:ORCE 1}
    \mPe
    \leq
        \max_{k \in \T}
         e^{-N \tilde{E}_{\mathrm{r},k}\left(R_{k}\right)}.
\end{align}
Using \eqref{eq:ORCE 1}, we define the bottleneck error exponent
of the TWRC as follows.

\begin{definition}[Bottleneck Error Exponent] \label{def: MinimalRCE}

For a TWRC with the terminal $\mathrm{T}_{k \in \T}$ transmitting
a code $\left(e^{N R_k},N \right)$ of rate $R_k$, the bottleneck
error exponent at the information-exchange rate pair
$\left(R_1,R_2\right)$ is defined as
\begin{align} \label{eq:ORCE 2}
    E_{\mathrm{r}}^{\star}\left(R_1,R_2\right)
    &\triangleq
        \min_{k \in \T}
        \tilde{E}_{\mathrm{r},k}\left(R_{k}\right).
    \\[-0.5cm] \nonumber
\end{align}

\end{definition}

\begin{remark}

Using the RCEE of the link $\Link{k \in \T}$ in
Theorem~\ref{thm1}, we can readily obtain the bottleneck error
exponent $E_{\mathrm{r}}^{\star}\left(R_1,R_2\right)$. From
Definition~\ref{def: MinimalRCE}, we can see that the bottleneck
error exponent captures the behavior of the worst exponent decay
between the two links in the TWRC and reflects the reliability of
the information exchange at a rate pair $\left(R_1,R_2\right)$.
When the worst link is good enough, it means that the other link
must also be good. As a result, using \eqref{eq:ORCE 2} as an
information-exchange reliability measure, we can design a two-way
relay network such that both links can communicate reliably.
Besides the achievable rate region, we can also characterize the
bottleneck exponent plane from the set of all possible rate pairs.
This plane could provide us with further understanding of the
tradeoff between the rate pair $\left(R_{1}, R_{2}\right)$ and the
bottleneck error exponent (i.e., information-exchange
reliability).
\end{remark}


\section{Optimal Resource Allocation}

\subsection{Optimal Rate Allocation}

In the following, we present the optimal rate allocation that
maximizes the bottleneck error exponent
$E_{\mathrm{r}}^{\star}\left(R_1,R_2\right)$ under a sum-rate
constraint in the reliable information-exchange region
$\mathscr{R} = \left\{\left(R_1,R_2\right) : 0 \leq R_{1} \leq
\langle C_1 \rangle, 0 \leq R_{2} \leq \langle C_2
\rangle\right\}$. Mathematically, this rate allocation problem can
be formulated as follows:
\begin{align} \label{eq:rate 1}
    \mathcal{P}_1
    &=
        \left\{
          \begin{array}{ll}
            \displaystyle \max_{R_1,R_2}
                            &\quad
                                E_{\mathrm{r}}^{\star}\left(R_1,R_2\right)
            \\
            \hbox{s.t.} & \quad R_1 + R_2 = \SR
            \\
            &\quad 0 \leq R_1 \leq \langle C_1 \rangle, ~0 \leq R_2 \leq \langle C_2 \rangle
          \end{array}
        \right.
\end{align}
which can be solved by the following theorem.


\begin{theorem}\label{Thm OPRates}

Let $\mathscr{C}$ and $\mathscr{L}$ be the curve and straight line
in $\mathscr{R}$ such that
\begin{align*}
    \mathscr{C}
    &=
        \left\{
            \left(R_1,R_2\right) \in \mathscr{R}:
                \tilde{E}_{\mathrm{r},1}\left(R_1\right)
                =
                \tilde{E}_{\mathrm{r},2}\left(R_2\right)
        \right\}
    \\
    \mathscr{L}
    &=
        \left\{
            \left(R_1,R_2\right) \in \mathscr{R} :
                R_1+R_2=\SR
        \right\}.
\end{align*}
Then, the optimal solution $\left(R_1,R_2\right)_\mathrm{opt}$ of
the rate allocation problem $\mathcal{P}_1$ for the sum rate $\SR
\geq \left|\Ro{1}-\Ro{2}\right|$ is the intersection point of the
rate-pair curve $\mathscr{C}$ and straight line $\mathscr{L}$. In
particular, we have
\begin{align}   \label{eq:Thm OPRates}
    \left(R_1,R_2\right)_\mathrm{opt}
    =
        \begin{cases}
            \left(
                \dfrac{\SR+\Ro{1}-\Ro{2}}{2},
                \dfrac{\SR-\Ro{1}+\Ro{2}}{2}
            \right)
            &
            \text{for~~} \left|\Ro{1}-\Ro{2}\right| \leq \SR \leq
            \SR_\mathrm{d}^\star
            \\
            \,\left(\SR,0\right)
            &
            \text{for~~} \SR < \Ro{1}-\Ro{2}, \Ro{1} > \Ro{2}
            \\
            \,\left(0,\SR\right)
            &
            \text{for~~} \SR < \Ro{2}-\Ro{1}, \Ro{1} < \Ro{2}
        \end{cases}
\end{align}
where $\SR_\mathrm{d}^\star$ is the \emph{decisive sum rate} given
by
\begin{align} \label{eq:thm2}
    \SR_\mathrm{d}^\star
    =
        \min\left\{
            2\Rcr{1}-\Ro{1}+\Ro{2},
            2\Rcr{2}+\Ro{1}-\Ro{2}
        \right\}.
\end{align}

\begin{proof}
See Appendix~\ref{sec:proof:Thm OPRates}.
\end{proof}

\end{theorem}

\begin{remark}

For the sum rate $\SR > \SR_\mathrm{d}^\star$, we can determine
the \emph{quasi-optimal} rate pair as
\begin{align}   \label{eq:thm2:remark}
    \left(R_{1},R_{2}\right)_\mathrm{opt}
    \approx
        \begin{cases}
            \left(
                \dfrac{\SR+\Ro{1}-\Ro{2}}{2},
                \dfrac{\SR-\Ro{1}+\Ro{2}}{2}
            \right)
            &
            \text{for~~} \SR_\mathrm{d}^\star < \SR \leq
            \grave{\SR}_\mathrm{d}^\star
            \\
            \,\left(\SR-\langle C_2 \rangle,\langle C_2 \rangle\right)
            &
            \text{for~~} \SR >\grave{\SR}_\mathrm{d}^\star, \langle C_1 \rangle > \langle C_2 \rangle
            \\
            \,\left(\langle C_1 \rangle,\SR-\langle C_1 \rangle\right)
            &
            \text{for~~} \SR >\grave{\SR}_\mathrm{d}^\star, \langle C_1 \rangle < \langle C_2 \rangle
        \end{cases}
\end{align}
where $\grave{\SR}_\mathrm{d}^\star = \min\left\{ 2\langle C_1
\rangle-\Ro{1}+\Ro{2}, 2\langle C_2 \rangle+\Ro{1}-\Ro{2}
\right\}$.\footnote{
        The numerical example
        in Section~\ref{sec:NR} will show that this quasi-optimal rate
        pair well approximates the optimal
        one for the sum rate $\SR > \SR_\mathrm{d}^\star$.
        }
Therefore, with knowing the capacity and cutoff rate of each link
in the TWRC, we can determine the optimal rate pair
$\left(R_{1},R_{2}\right)_\mathrm{opt}$ that maximizes the
reliability of information exchange at the sum rate
$\SR$---exactly for $\SR \leq \SR_\mathrm{d}^\star$ using
\eqref{eq:Thm OPRates}, and approximately for $\SR >
\SR_\mathrm{d}^\star$ using \eqref{eq:thm2:remark}.

\end{remark}

\subsection{Optimal Power Allocation}

In this subsection, we present the optimal power allocation that
maximizes the bottleneck error exponent
$E_{\mathrm{r}}^{\star}\left(R_{1},R_{2}\right)$ at a rate pair
$\left(R_{1}, R_{2}\right)$. In the presence of perfect global
CSI, for fixed $\rho$ and $\left(R_{1}, R_{2}\right)$, we are
maximizing the instantaneous bottleneck exponent over
$\B{p}=\left[p_1 ~ p_2 \right]^{T}$ for each fading state, i.e.,
before averaging with respect to fading. Mathematically, we can
formulate this optimization problem as follows:
\begin{align} \label{eq: OPA 1}
    \mathcal{P}_2
    &=
        \left\{%
            \begin{array}{ll}
                 \displaystyle \max_{\B{p}} &\quad E_{\mathrm{r}}^{\mathrm{int}}\left(\B{p},\rho,R_1,R_2\right)
                 \\
                 \hbox{s.t.} &\quad p_1 + p_2 \leq P
                 \\
                 &\quad p_1 \geq 0, ~p_2 \geq 0
            \end{array}%
        \right.
\end{align}
where the subscript ``int'' denotes an instantaneous value and
\begin{align}
    E_{\mathrm{r}}^{\mathrm{int}}\left(\B{p},\rho,R_1,R_2\right)
    &=
        \min_{k \in \T}
        E_{\mathrm{r},k}^{\mathrm{int}}\left(\B{p},\rho, R_k\right)
    \\
    E_{\mathrm{r},k}^{\mathrm{int}}\left(\B{p},\rho,R_k\right)
    &\triangleq
        - \ln
        \left[
            1 + \frac{1}{1+\rho}
                \frac{p_k \PR \alpha_1 \alpha_2}{p_k \alpha_k + \left(\PR + p_1 p_2/p_k\right) \alpha_1\alpha_2/\alpha_k + 1}
        \right]^{-\rho}
        - 2 \rho R_k.
\end{align}
With the optimizing $\B{p}_\mathrm{opt}$ obtained by solving the
problem \eqref{eq: OPA 1}, we can find the bottleneck error
exponent with optimal power allocation as follows:
\begin{align} \label{eq:OPA}
    E_{\mathrm{r}}^{\star}\left(R_{1},R_{2}\right)
    =
        \EXs{\alpha_1,\alpha_2}{
            \max_{\rho \in \left[0,1\right]}
                E_{\mathrm{r}}^{\mathrm{int}}\left(\B{p}_\mathrm{opt},\rho,R_1,R_2\right)
        }.
\end{align}
Since $p_1$ and $p_2$ are positive, we can define $\psi_{k \in \T}
\triangleq \sqrt{p_k}$ and $\B{\psi} = \left[\psi_1 ~ \psi_2
\right]^{T}$ without loss of optimality. With this change of
variables, we can transform the optimization problem in \eqref{eq:
OPA 1} into a quasi-concave program.\footnote{
             Let $\mathcal{S}$ be a convex subset of
             $\mathbb{R}^{N}$. A function $f : \mathcal{S} \rightarrow
            \mathbb{R}$ is said to be quasi-convex if and only if its
            lower-level sets $L\left(f,a\right) = \{\B{x} \in
            \mathcal{S} : f\left(\B{x}\right) \leq a\}$ are convex sets for
            every $a \in \mathbb{R}$. Similarly, $f$ is said to be
            quasi-concave if and only if its upper-level sets
            $U\left(f,a\right) = \{\B{x} \in
            \mathcal{S} : f\left(\B{x}\right) \geq a\}$ are convex sets for every $a \in
            \mathbb{R}$.
            }
%

\begin{theorem}\label{thm_2}

For fixed $\rho$ and rate pair $\left(R_{1}, R_{2}\right)$, the
function
$E_{\mathrm{r}}^{\mathrm{int}}\left(\B{p},\rho,R_1,R_2\right)$ is
quasi-concave and the program $\mathcal{P}_2$ is quasi-concave.

\begin{proof}
See Appendix~\ref{sec:proof:thm2}.
\end{proof}

\end{theorem}

It is well known that we can solve quasi-convex optimization
problems efficiently through a sequence of convex feasibility
problems using the bisection method\cite{BV:04:Book}.\footnote{
            Note that the program
            $\mathcal{P}_2$ is always feasible as long as $P > 0$.
            }
We formalize it in the following corollary.

\begin{corollary}\label{lemma3}

The program $\mathcal{P}_2$ can be solved numerically using the
bisection method:

\begin{description}
  \item[0.] Initialize $t_{\mathrm{min}}$ and $t_{\mathrm{max}}$,
  where $t_{\mathrm{min}}$ and $t_{\mathrm{max}}$ define a range of relevant
  values of $E_{\mathrm{r}}^{\mathrm{int}}\left(\B{p},\rho,R_1,R_2\right)$, and set the
  tolerance $\varepsilon \in \mathbb{R}_{++}$.
  \item[1.] Solve the convex feasibility program
  $\mathcal{P}_{\mathrm{socp}}\left(t\right)$ in (\ref{eq:lemma3_1})
  by fixing $t = \left(t_{\mathrm{max}} + t_{\mathrm{min}}\right)/2$.
  \item[2.] If $\mathcal{S}\left(t\right) = \emptyset$, then set $t_{\mathrm{max}} = t$ else set $t_{\mathrm{min}} = t$.
  \item[3.] Stop if the gap $\left(t_{\mathrm{max}} - t_{\mathrm{min}}\right)$
  is less than the tolerance $\varepsilon$. Go to Step 1 otherwise.
  \item[4.] Output $\B{\psi}_{\mathrm{opt}}$ obtained from solving
  $\mathcal{P}_{\mathrm{socp}}\left(t\right)$ in Step 1.
\end{description}
where the convex feasibility program can be written in SOCP form
as \cite{LVBL:98:LAA}
\begin{align}\label{eq:lemma3_1}
\begin{array}{ll}
    \mathcal{P}_{\mathrm{socp}}\left(t\right) :
                        \hbox{find} &\quad \B{\psi}
                        \\
            ~~~~~~~~~~~~~\hbox{s.t.} &\quad \B{\psi} \in \mathcal{S}\left(t\right)
\end{array}
\end{align}
with the set $\mathcal{S}\left(t\right)$ given by
\begin{salign}\label{eq:lemma3_2}
    \mathcal{S}\left(t\right)
    & =
        \left\{
            \B{\psi} \in \mathbb{R}_{+}^{2}
            :
            \left[
                \begin{array}{c}
                    \B{\psi}^{T}\B{e}_{1}/\sqrt{v_1}
                    \\
                    \left(
                        \begin{array}{c}
                             \B{A} \B{\psi}
                             \\
                             \sqrt{1 + \PR \alpha_2}
                        \end{array}
                    \right)
                \end{array}
            \right]
                \succeq_{\mathcal{K}} 0,
            ~\left[
                \begin{array}{c}
                    \B{\psi}^{T}\B{e}_{2}/\sqrt{v_2}
                    \\
                    \left(
                        \begin{array}{c}
                            \B{A} \B{\psi} \\
                            \sqrt{1 + \PR \alpha_1}
                        \end{array}
                    \right)
                \end{array}
            \right]
                \succeq_{\mathcal{K}} 0,
            ~\left[
                \begin{array}{c}
                    \sqrt{P}
                     \\
                     \B{\psi}
                \end{array}
            \right]
                \succeq_{\mathcal{K}} 0 \right\}
\end{salign}
where $\B{A}$ and $v_{k \in \T}$ are defined by \eqref{eq:A} and
\eqref{eq:v} in Appendix~\ref{sec:proof:thm2}, respectively.

\begin{proof}
It follows directly from the proof of Theorem~\ref{thm_2} and
\cite{LVBL:98:LAA} that we can represent the convex constraints in
the set $\mathcal{S}\left(t\right)$ in terms of SOC constraints.
\end{proof}

\end{corollary}

\begin{remark}
It is important that we initialize an interval that contains the
optimal solution. In our case, we can always let
$t_{\mathrm{min}}$ correspond to the uniform power allocation and
we only need to choose $t_{\mathrm{max}}$ appropriately.
\end{remark}

\section{Numerical Results} \label{sec:NR}

In this section, we provide some numerical results to illustrate
our analysis. In all examples, we choose $\Omega_{1}=0.5$,
$\Omega_{2}=2$, $p_\mathrm{R}=P$, and define $\mathsf{SNR}
\triangleq P/N_0$. Without power allocation, we further consider
equal power allocation between two terminals $\mathrm{T}_{k \in
\T}$, namely, $p_{1}=p_{2}=P/2$.\footnote{
        For one-way relaying, the RCEE, capacity, and cutoff rate
        are symmetric and equal for two links in the case of equal
        power allocation.
        }

\subsection{Random Coding Error Exponent}

\begin{figure}[t]
    \centerline{\includegraphics[width=0.75\textwidth]{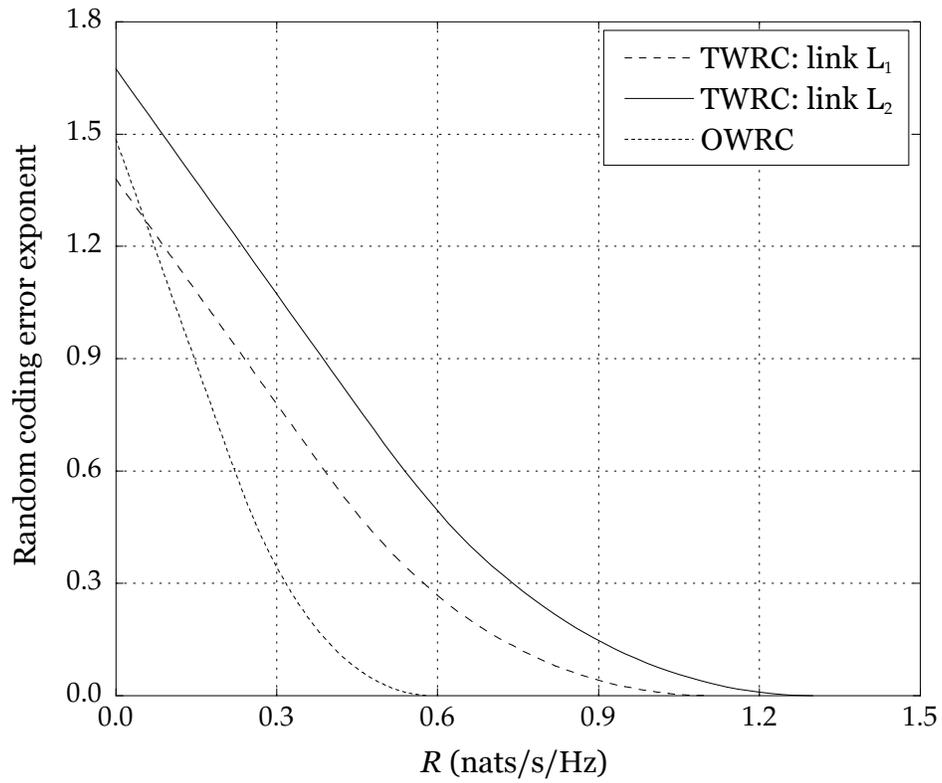}}
    \caption{Random coding error exponent for the link $\Link{k
            \in \T}$ of the TWRC and OWRC with ideal/hypothetical AF relaying.
            $\Omega_{1}=0.5$, $\Omega_{2}=2$, and $\mathsf{SNR}=20$ dB.}
    \label{fig:1}
\end{figure}

To ascertain the effectiveness of two-way relaying in terms of the
error exponent, Fig.~\ref{fig:1} shows the RCEE for the link
$\Link{k \in \T}$ of the TWRC and OWRC with ideal/hypothetical AF
relaying at $\mathsf{SNR}=20$ dB. To calculate the RCEE, we use
Theorem~\ref{thm1} for two-way relaying, whereas we modify
Theorem~\ref{thm1} for one-way relaying in such a manner as
described in Remark~\ref{remark:2}. We can see from the figure
that the link $\Link{2}$ of the TWRC shows better exponent
behavior than the link $\Link{1}$ at every rate $R$ due to the
fact that $\Omega_2>\Omega_1$. In the regime below the critical
rate, the exponent of the TWRC decreases with the rate twice as
slow as in the OWRC and hence, we require to increase the codeword
length slowly with two-way relaying to achieve the same level of
reliable information exchange as the rate increases. This is due
to the spectral efficiency of two-way relaying that requires only
half the time duration of one-way relaying to exchange the
information.

\subsection{Capacity and Cutoff Rate}

\begin{figure}[t]
    \centerline{\includegraphics[width=0.75\textwidth]{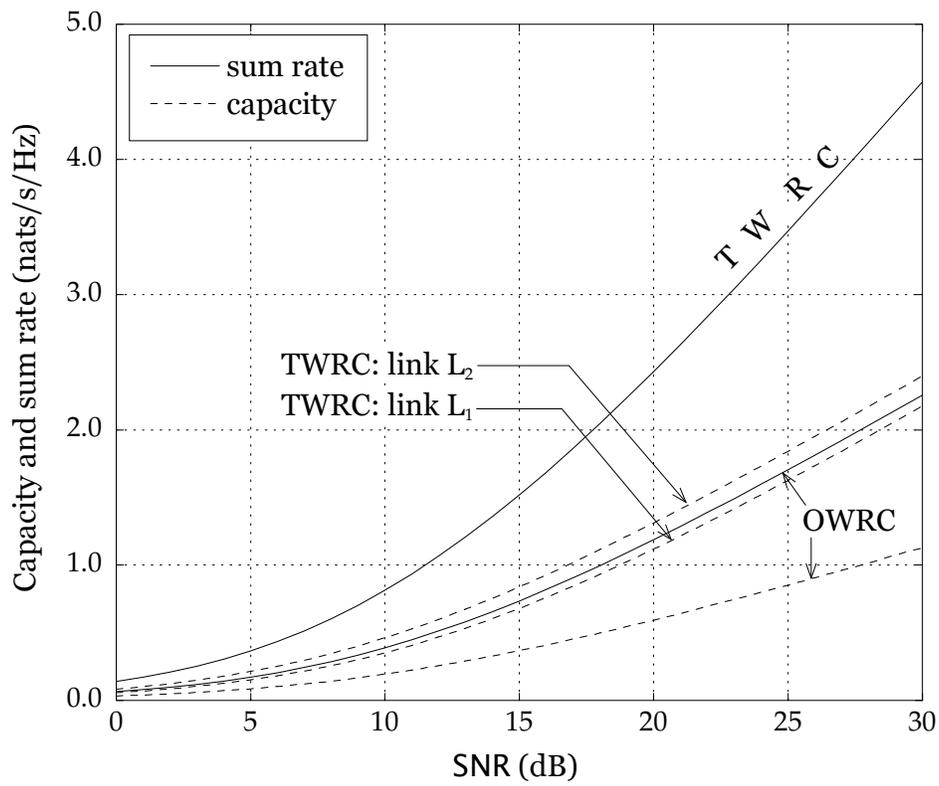}}
    \caption{Capacity and achievable sum rate versus $\mathsf{SNR}$ for the link $\Link{k
            \in \T}$ of the TWRC and OWRC with ideal/hypothetical AF relaying. $\Omega_{1}=0.5$ and $\Omega_{2}=2$.}
    \label{fig:2}
\end{figure}

\begin{figure}[t]
    \centerline{\includegraphics[width=0.75\textwidth]{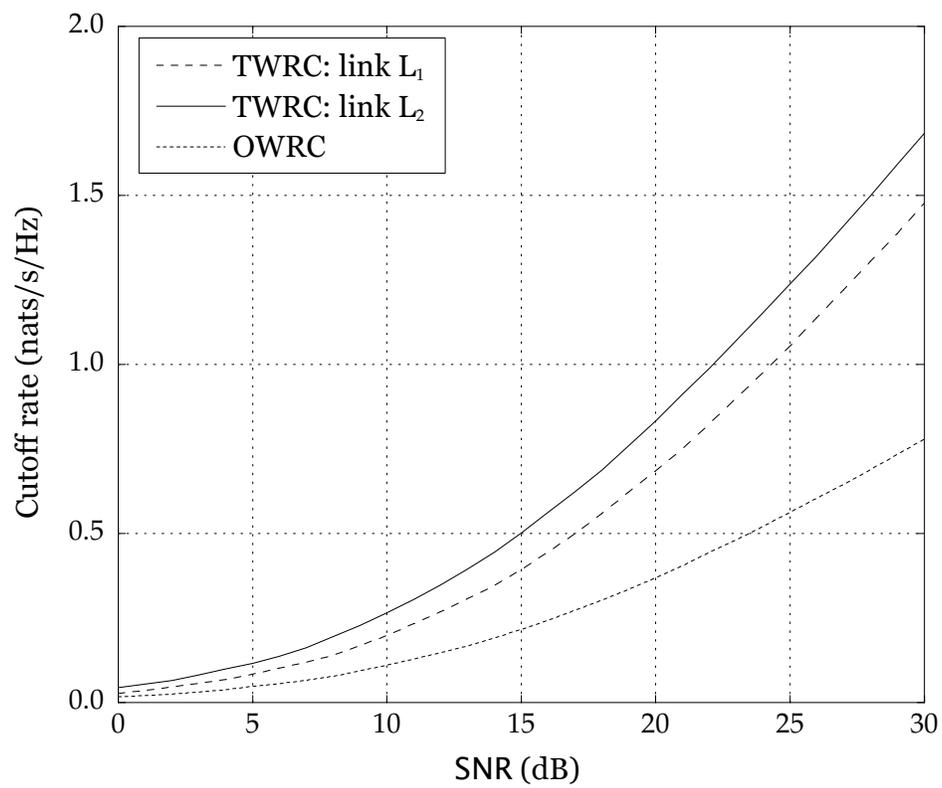}}
    \caption{Cutoff rate versus $\mathsf{SNR}$ for the link $\Link{k
            \in \T}$ of the TWRC and OWRC with ideal/hypothetical AF relaying. $\Omega_{1}=0.5$ and $\Omega_{2}=2$.}
    \label{fig:3}
\end{figure}

Figs.~\ref{fig:2} and \ref{fig:3} demonstrate the effectiveness of
two-way relaying on the achievable rates, where the capacity (or
achievable sum rate) and cutoff rate versus $\mathsf{SNR}$ are
depicted for the link $\Link{k \in \T}$ of the TWRC and OWRC with
ideal/hypothetical AF relaying, respectively. We can see from the
figures that the slopes of the capacity, achievable sum rate, and
cutoff rate curves at high $\mathsf{SNR}$ are twice as large in
the TWRC as in the OWRC due to again the fact that two-way
relaying for the information exchange can reduce the spectral
efficiency loss of half-duplex signaling by half in the TWRC.
Hence, as can be seen in Fig.~\ref{fig:2}, the high-$\mathsf{SNR}$
slope of the capacity for the link $\Link{k \in \T}$ of the TWRC
is identical to that of the achievable sum rate in the OWRC.

\subsection{Bottleneck Error Exponent}

\begin{figure}[t]
    \centerline{\includegraphics[width=0.75\textwidth]{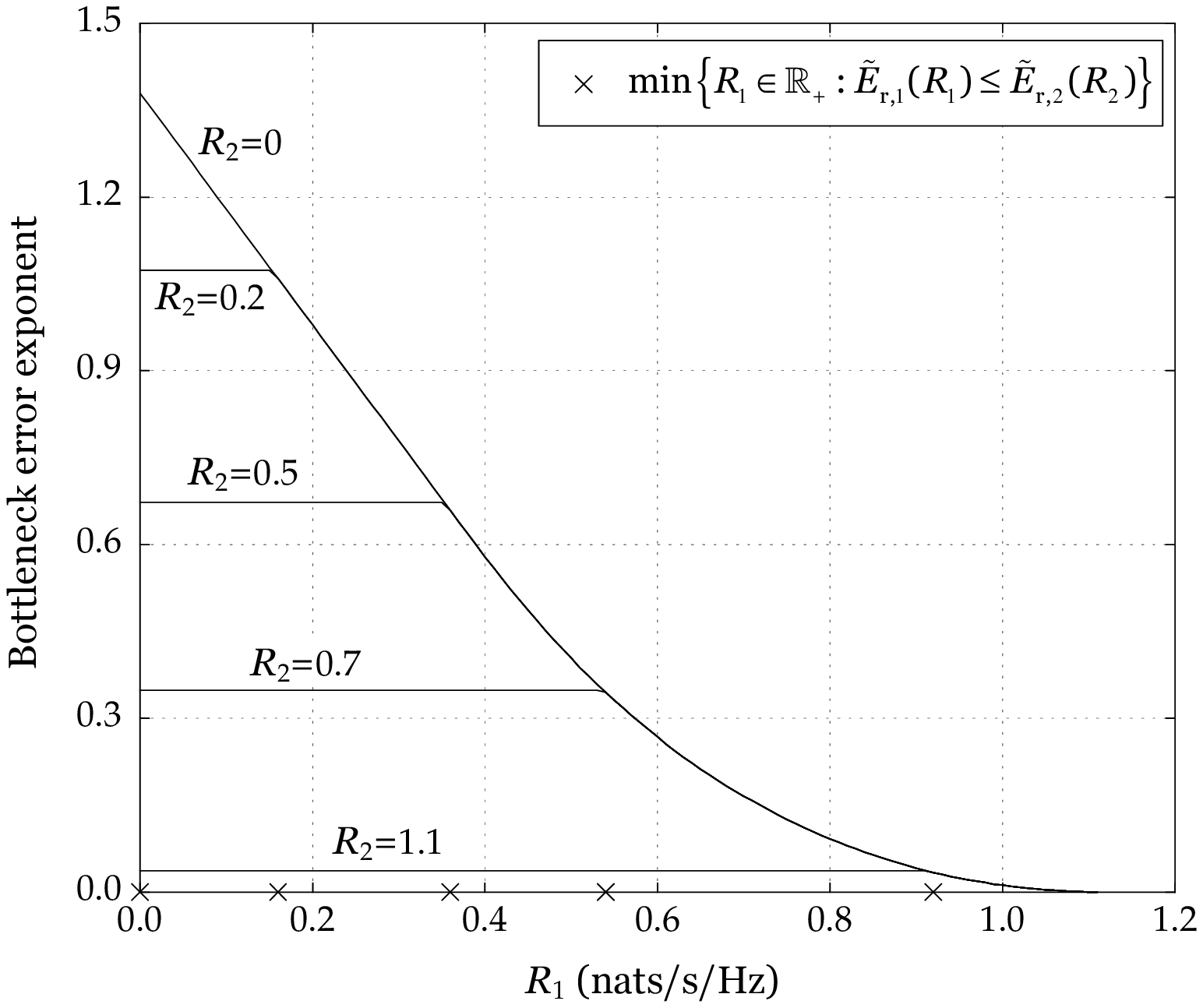}}
    \caption{Bottleneck error exponent $E_{\mathrm{r}}^{\star}\left(R_1,R_2\right)$ versus $R_1$
            for the TWRC with ideal/hypothetical AF relaying at $R_2=0$, $0.2$, $0.5$, $0.7$, and $1.1$ nats/s/Hz.
            $\Omega_{1}=0.5$, $\Omega_{2}=2$, and $\mathsf{SNR}=20$ dB. The values of $\min\bigl\{R_1 \in \mathbb{R}_+ :
            \tilde{E}_{\mathrm{r},1}\left(R_1\right) \leq
            \tilde{E}_{\mathrm{r},2}\left(R_2\right)\bigr\}$ are equal to $0$, $0.16$, $0.36$, $0.54$, $0.92$ nats/s/Hz for
            $R_2=0$, $0.2$, $0.5$, $0.7$, and $1.1$ nats/s/Hz, respectively (indicated by the cross marks). }
    \label{fig:RCE_R1}
\end{figure}

To demonstrate the tradeoff between the rate pair and
information-exchange reliability in the TWRC,
Fig.~\ref{fig:RCE_R1} shows the bottleneck error exponent
$E_{\mathrm{r}}^{\star}\left(R_1,R_2\right)$ versus $R_1$ for the
TWRC with ideal/hypothetical AF relaying at $\mathsf{SNR}=20$ dB
when $R_2=0$, $0.2$, $0.5$, $0.7$, and $1.1$. For fixed $R_2$, the
bottleneck exponent $E_{\mathrm{r}}^{\star}\left(R_1,R_2\right)$
as a function of $R_1$ behaves identically to
$\tilde{E}_{\mathrm{r},1}\left(R_1\right)$ at $R_1 \geq
R_1^\mathrm{min}= \min\bigl\{R_1 \in \mathbb{R}_+ :
\tilde{E}_{\mathrm{r},1}\left(R_1\right) \leq
\tilde{E}_{\mathrm{r},2}\left(R_2\right)\bigr\}$, whereas
$E_{\mathrm{r}}^{\star}\left(R_1,R_2\right)$ is limited to
$\tilde{E}_{\mathrm{r},1}\left(R_1^\mathrm{min}\right)$ for all
$R_1 \leq R_1^\mathrm{min}$. In this example, the values of
$R_1^\mathrm{min}$ are equal to $0$, $0.16$, $0.36$, $0.54$,
$0.92$ for $R_2=0$, $0.2$, $0.5$, $0.7$, and $1.1$, respectively.
As can be seen, the bad link in terms of the exponent is a
bottleneck that limits reliable information exchange, and the
bottleneck exponent at large $R_1$ or $R_2$ becomes small,
indicating the achievable reliability of information exchange
would be low at such rate pairs.

\subsection{Optimal Resource Allocation}

We now give application examples of the error exponent analysis
for the resource allocation in the TWRC.

\subsubsection{Optimal Rate Allocation}

\begin{figure}[t]
    \centerline{\includegraphics[width=0.8\textwidth]{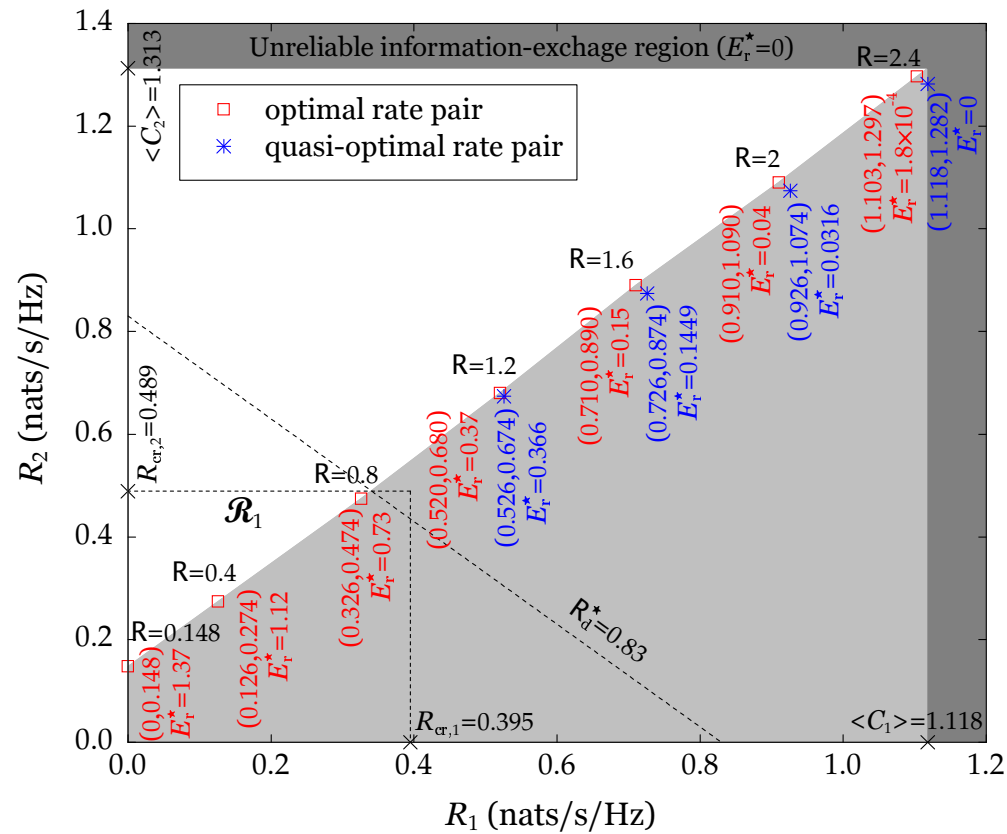}}
    \caption{Optimal rate pair $\left(R_1, R_2\right)_\mathrm{opt}$ that
            maximizes the bottleneck error exponent
            $E_{\mathrm{r}}^{\star}\left(R_1,R_2\right)$ for the
            TWRC with ideal/hypothetical AF relaying at sum
            rates $\SR=0.148$, $0.4$, $0.8$, $1.2$, $1.6$, $2.0$, and
            $2.4$ nats/s/Hz.
            $\Omega_{1}=0.5$, $\Omega_{2}=2$, and $\mathsf{SNR}=20$ dB.
            For $\SR>\SR_\mathrm{d}^\star=0.83$, the quasi-optimal rate
            pairs are also plotted for $\SR=1.2$, $1.6$, $2.0$, and $2.4$
            nats/s/Hz.
            }
    \label{fig:6}
\end{figure}

Fig.~\ref{fig:6} shows the optimal rate pair $\left(R_1,
R_2\right)_\mathrm{opt}$ that maximizes the bottleneck error
exponent $E_{\mathrm{r}}^{\star}\left(R_1,R_2\right)$ under a sum
rate constraint for the TWRC with ideal/hypothetical AF relaying
at $\mathsf{SNR}=20$ dB. The quasi-optimal rate pairs are also
plotted for the sum rate $\SR>\SR_\mathrm{d}^\star$. The optimal
and quasi-optimal rate pairs are determined using Theorem~\ref{Thm
OPRates} and \eqref{eq:thm2:remark}, respectively. The optimal
rate pairs $\left(R_1, R_2\right)_\mathrm{opt}$ at the sum rates
$\SR=0.148$, $0.4$, $0.8$, $1.2$, $1.6$, $2.0$, and $2.4$ are
$\left(0,0.148\right)$, $\left(0.126,0.274\right)$,
$\left(0.326,0.474\right)$, $\left(0.520,0.680\right)$,
$\left(0.710,0.890\right)$, $\left(0.910,1.090\right)$, and
$\left(1.103,1.297\right)$, attaining the maximum
$E_{\mathrm{r}}^{\star}\left(R_1,R_2\right)$ equal to $1.37$,
$1.12$, $0.73$, $0.37$, $0.15$, $0.04$, and $1.8\times 10^{-4}$,
respectively. For $\SR>\SR_\mathrm{d}^\star=0.83$, the
quasi-optimal rate pairs at the sum rates $\SR=1.2$, $1.6$, $2.0$,
and $2.431$ are $\left(0.526,0.674\right)$,
$\left(0.726,0.874\right)$, $\left(0.926,1.074\right)$, and
$\left(1.118,1.282\right)$, attaining
$E_{\mathrm{r}}^{\star}\left(R_1,R_2\right)$ equal to $0.366$,
$0.1449$, $0.0316$, and $0$, respectively. We can see that the
quasi-optimal rate pairs quite well approximate the optimal
$\left(R_1, R_2\right)_\mathrm{opt}$ for
$\SR>\SR_\mathrm{d}^\star$ and achieve the bottleneck exponents
very close to the maximum achievable
$E_{\mathrm{r}}^{\star}\left(R_1,R_2\right)$ at such sum rates. In
the figure, the region $\mathscr{R}$ can be divided by the optimal
rate curve into two rate-pair subregions in which each RCEE
$\tilde{E}_{\mathrm{r},k}\left(R_k\right)$ is dominant for the
bottleneck exponent $E_{\mathrm{r}}^{\star}\left(R_1,R_2\right)$:
for example, $\tilde{E}_{\mathrm{r},1}\left(R_1\right)$ is
dominant in the light gray subregion, i.e.,
$E_{\mathrm{r}}^{\star}\left(R_1,R_2\right)=
\tilde{E}_{\mathrm{r},1}\left(R_1\right)$.

\begin{figure}[t]
    \centerline{\includegraphics[width=0.75\textwidth]{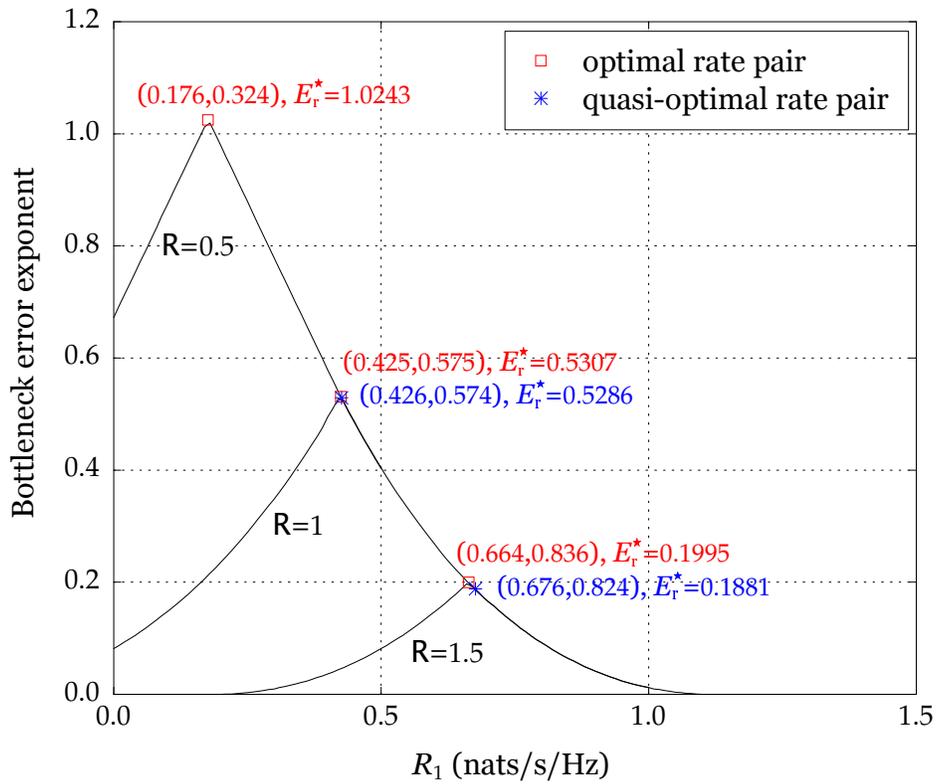}}
    \caption{Bottleneck error exponent $E_{\mathrm{r}}^{\star}\left(R_1,R_2\right)$
            versus $R_1$ for the TWRC with ideal/hypothetical AF relaying at sum rates
            $\SR = 0.5$, $1$, and $1.5$ nats/s/Hz.
            $\Omega_{1}=0.5$, $\Omega_{2}=2$, and $\mathsf{SNR}=20$
            dB. The optimal rate pair $\left(R_1,
            R_2\right)_\mathrm{opt}$ for each sum rate and the quasi-optimal rate
            pairs for $\SR>\SR_\mathrm{d}^\star=0.83$ are also
            plotted.
            }
    \label{fig:7}
\end{figure}

The effectiveness of the optimal/quasi-optimal rate allocation in
maximizing the bottleneck error exponent can be further
ascertained by referring Fig.~\ref{fig:7}, where the bottleneck
error exponent $E_{\mathrm{r}}^{\star}\left(R_1,R_2\right)$ versus
$R_1$ is depicted for the TWRC with ideal/hypothetical AF relaying
at $\mathsf{SNR}=20$ dB when the sum rate $\SR=R_1+R_2$ is fixed
to $0.5$, $1$, and $1.5$, respectively. As can be seen from the
figure, the bottleneck exponent
$E_{\mathrm{r}}^{\star}\left(R_1,R_2\right)$ is unimodal as a
function of $R_1$ for fixed $\SR$, and its maximum is at the mode
of $R_1$ determined by Theorem~\ref{Thm OPRates} for each value of
$\SR$. We can also observe that the optimal/quasi-optimal rate
allocation is of significant benefit to the bottleneck exponent.
The optimal rate pairs $\left(R_1, R_2\right)_\mathrm{opt}$ at the
sum rates $\SR=0.5$, $1$, and $1.5$ are
$\left(0.176,0.324\right)$, $\left(0.425,0.575\right)$, and
$\left(0.664,0.836\right)$, attaining the maximum
$E_{\mathrm{r}}^{\star}\left(R_1,R_2\right)$ equal to $1.0243$,
$0.5307$, and $0.1995$, respectively. For
$\SR>\SR_\mathrm{d}^\star=0.83$, the quasi-optimal rate pairs at
the sum rates $\SR=1$ and $1.5$ are $\left(0.426,0.574\right)$ and
$\left(0.676,0.824\right)$, attaining
$E_{\mathrm{r}}^{\star}\left(R_1,R_2\right)$ equal to $0.5286$ and
$0.1881$, respectively. We can see again that the quasi-optimal
rate pairs quite well approximate the optimal ones for
$\SR>\SR_\mathrm{d}^\star$ with a negligible loss in the
bottleneck exponent.

\subsubsection{Optimal Power Allocation}

\begin{figure}[t]
    \centerline{\includegraphics[width=0.75\textwidth]{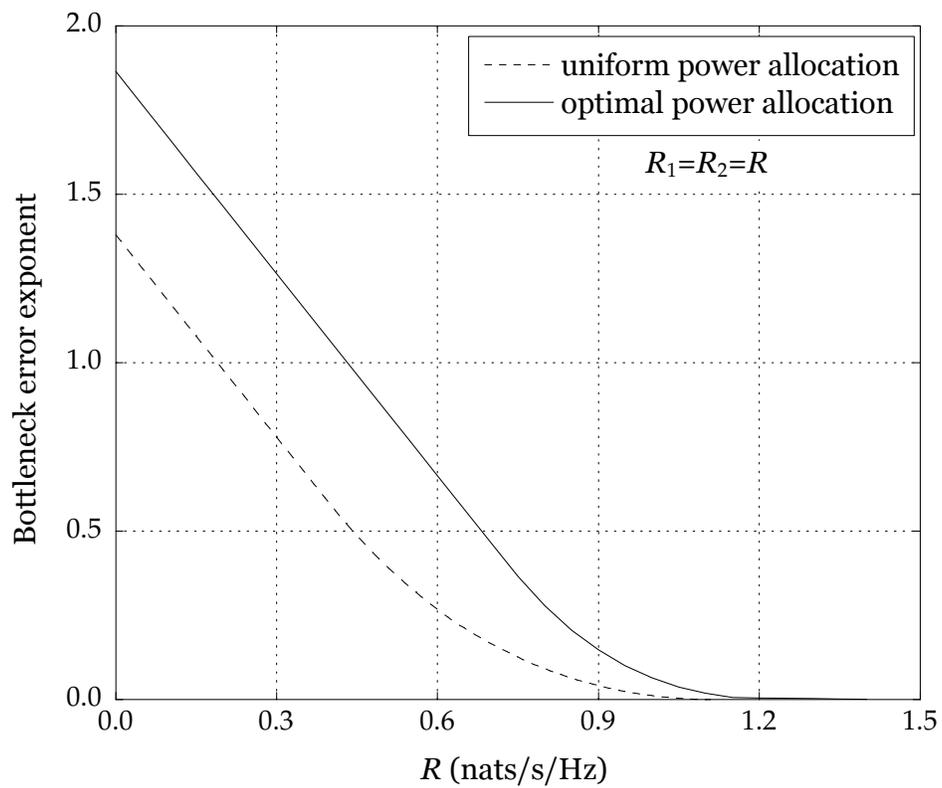}}
    \caption{Bottleneck error exponent $E_{\mathrm{r}}^{\star}\left(R_1,R_2\right)$
            versus $R$ for the TWRC with ideal/hypothetical AF relaying with optimal
            and uniform power allocations. $R_1 = R_2=R$,
            $\Omega_{1}=0.5$, $\Omega_{2}=2$, and $\mathsf{SNR}=20$ dB.
            }
    \label{fig:8}
\end{figure}

Fig.~\ref{fig:8} shows the bottleneck error exponent
$E_{\mathrm{r}}^{\star}\left(R_1,R_2\right)$ versus $R=R_1=R_2$
for the TWRC with ideal/hypothetical AF relaying under optimal and
uniform power allocations at $\mathsf{SNR}= 20$ dB. To determine
$E_{\mathrm{r}}^{\star}\left(R_1,R_2\right)$ in \eqref{eq:OPA}, we
first find the optimal power allocation $\B{p}_\mathrm{opt}$ using
Corollary~\ref{lemma3} maximize
$E_{\mathrm{r}}^{\mathrm{int}}\left(\B{p}_\mathrm{opt},\rho,R_1,R_2\right)$
over $\rho$ using the method given in
\cite[Section~2.2.4]{Ahm:97:PhD}, and then successively perform
the expectation of $\max_{\rho \in \left[0,1\right]}
E_{\mathrm{r}}^{\mathrm{int}}\left(\B{p}_\mathrm{opt},\rho,R_1,R_2\right)$
with respect to $\alpha_{k \in \T}$ by the Monte Carlo method.
Compared with the uniform power allocation, we can see that the
optimal power allocation significantly improves the bottleneck
error exponent.

\section{Conclusions}

In this paper, we have derived Gallager's random coding exponent
to analyze the fundamental tradeoff between the communication
reliability and transmission rate in AF two-way relay channels.
The exponent has been expressed in terms of the generalized Fox
$H$-function, from which the capacity and cutoff rate were also
deduced for the link of each direction in the TWRC. Using the
worst exponent decay between two links as the reliability measure
for the information exchange, we put forth the concept of the
bottleneck error exponent to effectively capture the tradeoff
between the rate pair of the two links and the
information-exchange reliability at such a rate pair for the
design of two-way relay networks such that both links can
communicate reliably. As its applications, we formulated the
optimal rate and power allocation problems that maximize the
bottleneck error exponent. Specifically, we presented the optimal
rate allocation under a sum-rate constraint and its simple
closed-form quasi-optimal solution that requires knowing only the
capacity and cutoff rate of each link. The optimal power
allocation under a total power constraint of the two terminals was
further determined in the presence of perfect global CSI by
solving the quasi-convex optimization problem.

\appendix

\subsection{Proof of Theorem~\ref{thm1}}    \label{sec:proof:thm1}

Let $V_k=p_k \alpha_k$ and $W_k=\left(p_\mathrm{R}+p_1 p_2 /p_k
\right) \alpha_1 \alpha_2/ \alpha_k$. Then, $V_k \sim
\Erv{\lambda_k}$, $W_k \sim \Erv{\mu_k}$, and
\begin{align}\label{eq:proofthm1 2}
    \gamma^{\mathrm{ub}}_{k \in \T}
    =
        \frac{\eta_{k} V_k W_k}{V_k + W_k}.
\end{align}
%
%
%
%
Using the probability density function (PDF) of the Harmonic mean
of the two exponential random variables \cite{HA:03:WCOM} and the
transformation $\PDF{Y}{y}=\frac{1}{|a|}\PDF{X}{y/a} $ where
$Y=aX$, we obtain the PDF of $\gamma^{\mathrm{ub}}_{k}$ as
\begin{align}\label{eq:proofthm1 3}
    \PDF{\gamma^{\mathrm{ub}}_{k}}{\gamma}
    &=
         \frac{4}{\eta_{k}^2}
         \lambda_k \mu_k
         \gamma
         e^{-\frac{\left(\lambda_k +\mu_k\right)\gamma}{\eta_{k}}}
         K_0\left(\frac{2 \gamma \sqrt{\lambda_k\mu_k}}{\eta_{k}}\right)
    \nonumber \\
    &\hspace{0.5cm}
        +
         \frac{2}{\eta_{k}^2}
         \left(\lambda_k+\mu_k\right)
         \sqrt{\lambda_k\mu_k}
         \gamma
         e^{-\frac{\left(\lambda_k + \mu_k\right)\gamma}{\eta_{k}}}
         K_1\left(\frac{2 \gamma \sqrt{\lambda_k\mu_k}}{\eta_{k}}\right),
    \quad
    \gamma \geq 0
\end{align}
where $K_\nu\left(\cdot\right)$ is the $\nu$th order modified
Bessel function of the second kind whose integral representation
is given by \cite[eq.~(8.432.6)]{GR:00:Book}.

Using \eqref{eq:proofthm1 3}, we have
\begin{align}\label{eq:proofthm1 1}
    \tilde{E}_{0,k}\left(\rho\right)
    &=
        -\ln\left\{
        \int_0^\infty
            \left(
                    1 + \frac{\gamma}{1 + \rho}
            \right)^{-\rho}
            \PDF{\gamma^{\mathrm{ub}}_{k}}{\gamma}
            d\gamma
        \right\}.
\end{align}
Since it is obvious that $\tilde{E}_{0,k}\left(\rho\right)=0$ for
$\rho=0$, we define
\begin{align}\label{eq:proofthm1 4}
    \mathcal{I}\left(\rho\right)
    \triangleq
        \int_0^\infty
            x \left(1+a x\right)^{-\rho}
            e^{-bx}
            K_\nu\left(cx\right)
            dx
\end{align}
to find $\tilde{E}_{0,k}\left(\rho\right)$ in \eqref{eq:proofthm1
1} for $0< \rho \leq 1$. To evaluate the integral
$\mathcal{I}\left(\rho\right)$, we first express
$\left(1+ax\right)^{-\rho}$ and $e^{cx} K_\nu\left(cx\right)$ in
terms of the Fox $H$-functions with the help of
\cite[eqs.~(8.3.2.21), (8.4.2.5), and (8.4.23.5)]{PBM:90:Book:v3}
as follows:
\begin{salign}\label{eq:proofthm1 5}
    \left(1+ax\right)^{-\rho}
    &=
        \frac{1}{\Gamma\left(\rho\right)}
        \FoxH{1}{1}{1}{1}{a x}{\left(1-\rho,1\right)}{\left(0,1\right)}
    \\
        \label{eq:proofthm1 6}
    e^{cx} K_\nu\left(cx\right)
    &=
        \frac{\cos\left(\nu\pi\right)}{\sqrt{\pi}}
        \FoxH{2}{1}{1}{2}{2cx}{\left(1/2,1\right)}{\left(\nu,1\right), \left(-\nu,1\right)}
\end{salign}
where $H_{p,q}^{m,n}\left[\cdot\right]$ is the Fox $H$-function
\cite[eq.~(8.3.1.1)]{PBM:90:Book:v3}. Then, substituting
\eqref{eq:proofthm1 5} and \eqref{eq:proofthm1 6} into
\eqref{eq:proofthm1 4}, we have
\begin{salign}\label{eq:proofthm1 8}
    \mathcal{I}\left(\rho\right)
    &=
        \frac{\cos\left(\nu\pi\right)}{\sqrt{\pi}\Gamma\left(\rho\right)}
        \int_0^\infty
            x
            e^{-\left(b+c\right)x}
            \FoxH{1}{1}{1}{1}{a x}{\left(1-\rho,1\right)}{\left(0,1\right)}
            \FoxH{2}{1}{1}{2}{2cx}{\left(1/2,1\right)}{\left(\nu,1\right), \left(-\nu,1\right)}
            dx
    \nonumber \\
    &=
        \frac{\cos\left(\nu\pi\right)}{\sqrt{\pi}\Gamma\left(\rho\right)}
        \left(b+c\right)^{-2}
        \GFoxH{1,1,1,1,2}
            {1,\left(1:1\right),0,\left(1:2\right)}
            {\frac{a}{b+c}}
            {\frac{2c}{b+c}}
            {\left(2,1\right)}
            {\left(1-\rho,1\right);\left(1/2,1\right)}
            {\hspace{1.5cm}\text{---}}
            {\left(0,1\right);\left(\nu,1\right),\left(-\nu,1\right)}.
\end{salign}
where the last equality follows from
\cite[eq.~(2.6.2)]{MS:78:Book}. Finally, from \eqref{eq:proofthm1
3}--\eqref{eq:proofthm1 4} and \eqref{eq:proofthm1 8}, we get
\eqref{eq:thm1 3} and complete the proof.

\subsection{Proof of Corollary~\ref{coro1: capacity}}  \label{sec:proof:cor1}

It follows from Theorem~\ref{thm1} that
\begin{align}\label{eq:proofCorro 1}
    \langle C_{k}\rangle
    &=
        \frac{1}{2}
        \left.
        \left[
            \frac{\partial \tilde{E}_{0,k}\left(\rho\right)}{\partial \rho}
        \right]
        \right|_{\rho=0}
    \nonumber \\
    &=
     \frac{1}{2}
     \int_0^\infty
            \ln
            \left(
                1+\gamma
            \right)
                p_{\gamma^{\mathrm{ub}}_{k}}\left(\gamma\right)
                d\gamma.
\end{align}
Similar to the derivation of $\tilde{E}_{0,k}\left(\rho\right)$,
we first express $\ln\left(1+\gamma\right)$ in terms of the Fox
$H$-function with the help of \cite[eq.~(8.4.6.5)]{PBM:90:Book:v3}
as
\begin{salign}
    \ln\left(1+\gamma\right)
    =
        \FoxH{1}{2}{2}{2}{\gamma}{\left(1,1\right),\left(1,1\right)}{\left(1,1\right),\left(0,1\right)}.
\end{salign}
Then, again using \eqref{eq:proofthm1 6} and
\cite[eq.~(2.6.2)]{MS:78:Book}, we evaluate \eqref{eq:proofCorro
1} as \eqref{eq:coro1 1} and complete the proof.

\subsection{Proof of Theorem~\ref{Thm OPRates}} \label{sec:proof:Thm OPRates}

\begin{figure}[t]
    \centerline{\includegraphics[width=0.7\textwidth]{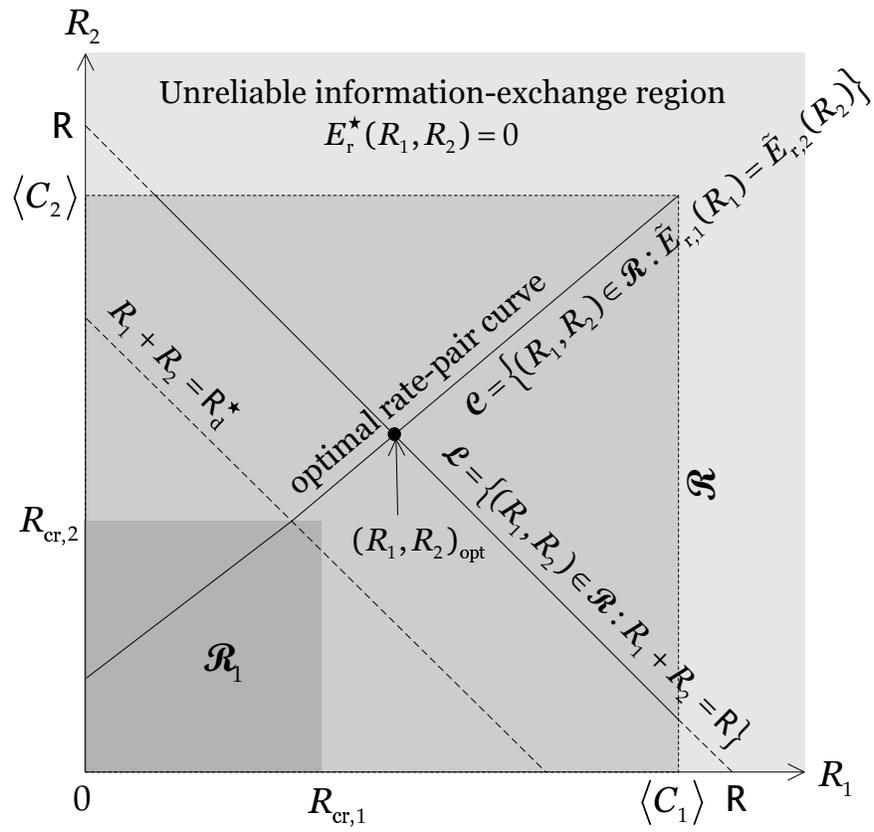}}
    \caption{Graphical interpretation of the optimal rate pair $\left(R_1, R_2\right)_\mathrm{opt}$.}
    \label{fig:9}
\end{figure}

Since the exponent $\tilde{E}_{\mathrm{r},k}\left(R_k\right)$ is a
monotonically decreasing function in $R_k$, it is obvious that for
any rate pair $\bigl(\grave{R}_{1},\grave{R}_{2}\bigr) \in
\mathscr{R}$ with the sum rate $\grave{R}_1+\grave{R}_2=\SR \geq
\left|\Ro{1}-\Ro{2}\right|$,
\begin{align}
    E_{\mathrm{r}}^{\star}\left(R_{1},R_{2}\right)
    \geq
        E_{\mathrm{r}}^{\star}\bigl(\grave{R}_{1},\grave{R}_{2}\bigr)
\end{align}
whenever $\left(R_1,R_2\right)$ is such that
$\tilde{E}_{\mathrm{r},1}\left(R_1\right) =
\tilde{E}_{\mathrm{r},2}\left(R_2\right)$ and $R_1+R_2=\SR$.
Therefore, the optimal solution
$\left(R_1,R_2\right)_\mathrm{opt}$ of the problem $\mathcal{P}_1$
for $\SR \geq \left|\Ro{1}-\Ro{2}\right|$ is uniquely given by
\begin{align} \label{eq:thm2:proof:R2}
    \left(R_{1},R_{2}\right)_\mathrm{opt}
    \in
        \left\{
            \left(R_1,R_2\right) \in \mathscr{R}:
                \tilde{E}_{\mathrm{r},1}\left(R_1\right)
                =
                \tilde{E}_{\mathrm{r},2}\left(R_2\right),~
                R_{1} + R_{2} = \SR
        \right\}.
\end{align}
Although, clearly, the optimization problem \eqref{eq:rate 1} is
mathematically challenging, it follows from
\eqref{eq:thm2:proof:R2} that the optimal solution
$\left(R_1,R_2\right)_\mathrm{opt}$ for $\SR \geq
\left|\Ro{1}-\Ro{2}\right|$ is the intersection point of the
rate-pair curve $\mathscr{C}$ and straight line $\mathscr{L}$, and
we can determine it graphically, as shown in Fig.~\ref{fig:9}.

Let
    $\mathscr{R}_1
    =
        \left\{
            \left(R_1,R_2\right) \in \mathscr{R}:
            0 \leq R_{1} \leq \Rcr{1},
            0 \leq R_{2} \leq \Rcr{2}
        \right\}$
and $\SR_\mathrm{d}^\star$ be the largest sum rate at which the
optimal solution $\left(R_1,R_2\right)_\mathrm{opt}$ of the
problem $\mathcal{P}_1$ belongs to the subregion $\mathscr{R}_1$.
When the rate is less than the critical rate, the optimal value of
$\rho$ is equal to $1$ and the RCEE for the link $\Link{k \in \T}$
of the TWRC can be written as
\begin{align} \label{eq:OptRate 2}
    \tilde{E}_{\mathrm{r},k}\left(R_k\right)
    &=
        \tilde{E}_{0,k}\left(1 \right) -  2 R_k  \nonumber
        \\
    &=
        2\left(\Ro{k}- R_k\right).
\end{align}
Therefore, for the sum rate $\SR \leq \SR_\mathrm{d}^\star$, the
problem $\mathcal{P}_1$ can be rewritten as
\begin{align} \label{eq:Proof OPRates 1}
    \mathcal{P}_1
    =
        \left\{
             \begin{array}{ll}
                     \displaystyle \max_{R_1, R_2}
                                    &\quad
                                    \min_{k \in \T}
                                            \left(\Ro{k}-R_k\right)
                     \\
                     \hbox{s.t.} &\quad R_1 + R_2 = \SR
                     \\
                     &\quad 0 \leq R_1 \leq \Rcr{1}, ~0 \leq R_2 \leq \Rcr{2}
\end{array}
        \right.
\end{align}
which is equivalent to
%
\begin{align} \label{eq:Proof OPRates 2}
\mathcal{P}_1 =
\left\{%
\begin{array}{ll}
    \displaystyle \max_{R_1} &\quad \min \left\{\Ro{1}-R_1, \Ro{2}-\SR + R_1 \right\} \\
    \hbox{s.t.} &\quad 0 \leq R_1 \leq \SR \leq \Rcr{1}+\Rcr{2}.
\end{array}
\right.
\end{align}
Without loss of generality, we assume $\Ro{2} \geq \Ro{1}$, and we
can consider two different cases as follows:

\begin{itemize}

\item

When $\SR \leq \Ro{2}-\Ro{1}$, we have $\Ro{2}-\SR+ R_{1} \geq
\Ro{1}-R_1$ and
\begin{align} \label{eq:Proof OPRates 3}
    &\min
        \left\{
           \Ro{1}-R_1,\Ro{2}-\SR + R_1
        \right\}
    \nonumber \\
    &\quad=
        \Ro{1}-R_1
    \nonumber \\
    &\quad \leq
        \Ro{1}.
\end{align}
Thus, in this case, the optimal rate pair is
\begin{align} \label{eq:Proof OPRates 4}
    \left(R_1,R_2\right)_\mathrm{opt}
    =
        \left(0,\SR\right).
\end{align}

\item

When $\SR \geq \Ro{2} - \Ro{1}$, we need to consider two
additional cases.

If $\Ro{1}-R_1 \geq \Ro{2}-\SR + R_1$ or $R_1 \leq
\frac{1}{2}\left(\SR -\Ro{2}+\Ro{1}\right)$, then
\begin{align} \label{eq:Proof OPRates 5}
    &\min
        \left\{
            \Ro{1}-R_1, \Ro{2}-\SR + R_1
        \right\}
    \nonumber \\
    &\quad=
        \Ro{2}-\SR + R_1
    \nonumber \\
    &\quad\leq
        \Ro{2}-\SR +
        \frac{\SR- \Ro{2}+\Ro{1}}{2}
    \nonumber \\
    &\quad=
        \frac{-\SR+\Ro{2}+\Ro{1}}{2}
\end{align}

Therefore, the optimal rate pair is given by
\begin{align} \label{eq:Proof OPRates 6}
    \left(R_1,R_2\right)_\mathrm{opt}
    =
        \left(
            \frac{\SR+\Ro{1}-\Ro{1}}{2},
            \frac{\SR-\Ro{1}+\Ro{1}}{2}
        \right).
\end{align}

If $\Ro{1}-R_1 \leq \Ro{2}-\SR + R_1$ or $R_1 \geq
\frac{1}{2}\left(\SR -\Ro{2}+\Ro{1}\right)$, then
\begin{align} \label{eq:Proof OPRates 7}
    &\min
        \left\{
            \Ro{1}-R_1, \Ro{2}-\SR + R_1
        \right\}
    \nonumber \\
    &\quad=
        \Ro{1}- R_1
    \nonumber \\
    &\quad\leq
        \Ro{1}-
        \frac{\SR- \Ro{2}+\Ro{1}}{2}
    \nonumber \\
    &\quad=
        \frac{-\SR+\Ro{2}+\Ro{1}}{2}
\end{align}
Therefore, the optimal rate pair is given by
\begin{align} \label{eq:Proof OPRates 8}
    \left(R_1,R_2\right)_\mathrm{opt}
    =
        \left(
            \frac{\SR+\Ro{1}-\Ro{1}}{2},
            \frac{\SR-\Ro{1}+\Ro{1}}{2}
        \right).
\end{align}
Since $\left(R_1,R_2\right)_\mathrm{opt}$ should belong to
$\mathscr{R}_1$, we can find the decisive sum rate
$\SR_\mathrm{d}^\star$ as \eqref{eq:thm2} from the fact that
\begin{align} \label{eq:Proof OPRates 9}
    \begin{cases}
        \dfrac{\SR+\Ro{1}-\Ro{1}}{2} \leq \Rcr{1}
        \\
        \dfrac{\SR-\Ro{1}+\Ro{1}}{2} \leq \Rcr{2}.
    \end{cases}
\end{align}

\end{itemize}

From \eqref{eq:Proof OPRates 4}, \eqref{eq:Proof OPRates 6}, and
\eqref{eq:Proof OPRates 8}, we arrive at the desired result
\eqref{eq:Thm OPRates}.

\subsection{Proof of Theorem~\ref{thm_2}} \label{sec:proof:thm2}

For any $t \in \mathbb{R}_{+}$, the upper-level set of
$E_{\mathrm{r},k}^{\mathrm{int}}\left(R_k\right)$ that belongs to
$\mathcal{S}$ is given by
\begin{salign}\label{eq:lemma2_proof1}
    U\left(E^{\mathrm{int}}_{\mathrm{r},k},t\right)
    &=
        \left\{
            \B{\psi} \in \mathbb{R}_{+}^{2} :
                - \ln
                    \left[
                        1 + \frac{1}{1+\rho}
                            \frac{\psi_k^2 \PR \alpha_1 \alpha_2}{\psi_k^2 \alpha_k + \left(\PR + \psi_1^2\psi_2^2/\psi_k^2\right) \alpha_1\alpha_2/\alpha_k + 1}
                    \right]
                        ^{- \rho} - 2 \rho R_k \geq t
        \right\} \nonumber
    \\
    & =
        \left\{
            \B{\psi} \in \mathbb{R}_{+}^{2} :
                \frac{\psi_k^2}{\psi_k^2 \alpha_k + \left(\PR + \psi_1^2\psi_2^2/\psi_k^2\right) \alpha_1\alpha_2/\alpha_k + 1} \geq v_k
        \right\} \nonumber
    \\
    & =
        \left\{
            \B{\psi} \in \mathbb{R}_{+}^{2} :
             \frac{\B{\psi}^{T}
             \B{e}_{k}}{\sqrt{v_k}} \geq \sqrt{1 +  \PR \alpha_1\alpha_2/\alpha_k + \|
                \B{A} \B{\psi} \|^2
        } \right\} \nonumber
     \\
    &=
        \left\{
            \B{\psi} \in \mathbb{R}_{+}^{2} :
                    \left[
                        \begin{array}{c}
                            \B{\psi}^{T} \B{e}_k / \sqrt{v_k}
                             \\
                            \left(
                                \begin{array}{c}
                                \B{A} \B{\psi}
                                \\
                                \sqrt{1 + \PR \alpha_1\alpha_2/\alpha_k}
                                \end{array}
                            \right)
                        \end{array}
                    \right]
                    \succeq_{\mathcal{K}} 0
        \right\}
\end{salign}
with
\begin{align}
            \label{eq:A}
    \B{A}
    &\triangleq
        \mathrm{diag}\left(\sqrt{\alpha_1},\sqrt{\alpha_2}\right)
    \\
            \label{eq:v}
    v_k
    &\triangleq
        \frac{\left(1 + \rho\right)}{\PR \alpha_1 \alpha_2}
        \left[\exp \left(\frac{t + 2\rho R_k}{\rho}\right) - 1 \right]
\end{align}
where $\succeq_{\mathcal{K}}$ denotes the generalized inequality
with respect to the second-order cone (SOC) $\mathcal{K}$
\cite{BV:04:Book} and $\B{e}_{k}$ is a standard basis vector with
a one at the $k$th element. It is clear that
$U\left(E^{\mathrm{int}}_{\mathrm{r},k},t\right)$ is a convex set
since it can be represented as an SOC. Since the upper-level set
$U\left(E^{\mathrm{int}}_{\mathrm{r},k},t\right)$ is convex for
every $t \in \mathbb{R}_{+}$,
$E^{\mathrm{int}}_{\mathrm{r},k}\left(\B{p},\rho,R_k\right)$ is,
thus, quasi-concave.\footnote{
            Note that a concave function is also
            quasi-concave.
            }

We now show that
$E_{\mathrm{r},k}^{\mathrm{int}}\left(\B{p},\rho,R_k\right)$ is
not concave by contradiction. Since the function
$\ln\left(\cdot\right)$ is a monotonic function, we simply need to
show that $f_k\left(\B{\psi}\right) = \frac{\psi_k^2}{\psi_k^2
\alpha_k + \left(\PR + \psi_1^2\psi_2^2/\psi_k^2\right)
\alpha_1\alpha_2/\alpha_k + 1}$ is not concave. We consider
$\B{\psi}_{a}$ and $\B{\psi}_b$ such that $\B{\psi}_a =
\zeta\B{e}_{k}$ and $\B{\psi}_{b} = \delta \zeta\B{e}_{k}$ for $0
\leq \zeta \leq \sqrt{P}$ and $0 < \delta < 1$. Clearly,
$\B{\psi}_a$ and $\B{\psi}_b$ are feasible solutions of
$\mathcal{P}_2$. For any $\lambda \in \left[0,1\right]$, we have
\begin{align}\label{eq:lemma2_proof2}
    f_k\left(\lambda\B{\psi}_a + \left(1 - \lambda\right)\B{\psi}_b\right)
    &=
    \left(\alpha_k +
            \frac{1 + \PR \alpha_1\alpha_2/\alpha_k}
                {\zeta^2
                    \left[
                        \lambda + \delta
                            \left(1 -\lambda\right)
                    \right]^2}
    \right)^{-1}
    \triangleq
    g_k\left(\zeta\right)
\end{align}
where $g_k\left(\zeta\right)$ is clearly convex in $\zeta$. Due to
convexity of $g_k\left(\zeta\right)$, the following inequality
must hold
\begin{align}\label{eq:lemma2_proof3}
    g_k\left(\lambda \zeta_a
    + \left(1 - \lambda\right) \zeta_b\right)
    \leq
    \lambda g_k\left(\zeta_a\right)
    + \left(1 -\lambda\right)g_k\left(\zeta_b\right).
\end{align}
Now, by letting $\zeta_a = \zeta / \left(\lambda + \delta \left(1
- \lambda\right)\right)$ and $\zeta_b = \delta \zeta /
\left(\lambda + \delta \left(1 - \lambda\right)\right)$, we can
rewrite \eqref{eq:lemma2_proof3} as
\begin{align}\label{eq:lemma2_proof4}
    f_k\left(\lambda \B{\psi}_a + \left(1 - \lambda\right)\B{\psi}_b\right)
    \leq
    \lambda f_k\left(\B{\psi}_a\right) + \left(1 -\lambda\right)f_k\left(\B{\psi}_b\right).
\end{align}
Thus, we have showed that there exist $\B{\psi}_a, \B{\psi}_b \in
\mathbb{R}_{+}^{2}$ and $\lambda \in \left[0,1\right]$ such that
\eqref{eq:lemma2_proof4} holds. By contradiction,
$f_k\left(\B{\psi}\right)$ is not a concave function on
$\mathbb{R}_{+}^{2}$. Therefore, it follows that
$E_{\mathrm{r},k}^{\mathrm{int}}\left(\B{p},\rho,R_k\right)$ is
also not concave.

Since the nonnegative weighted minimum of quasi-concave functions
is quasi-concave \cite{BV:04:Book},
$E_{\mathrm{r}}^{\mathrm{int}}\left(\B{p},\rho,R_1,R_2\right)$ is
also quasi-concave. Furthermore, $\mathcal{P}_2$ is a
quasi-concave optimization problem since the constraint set in
$\mathcal{P}_2$ is convex in $\B{\psi}$.

\bibliographystyle{IEEEtran}
\bibliography{IEEEabrv,RCEE-Two-Way-arXiv-v2}

\begin{thebibliography}{10}
\providecommand{\url}[1]{#1}
\csname url@rmstyle\endcsname
\providecommand{\newblock}{\relax}
\providecommand{\bibinfo}[2]{#2}
\providecommand\BIBentrySTDinterwordspacing{\spaceskip=0pt\relax}
\providecommand\BIBentryALTinterwordstretchfactor{4}
\providecommand\BIBentryALTinterwordspacing{\spaceskip=\fontdimen2\font plus
\BIBentryALTinterwordstretchfactor\fontdimen3\font minus
  \fontdimen4\font\relax}
\providecommand\BIBforeignlanguage[2]{{%
\expandafter\ifx\csname l@#1\endcsname\relax
\typeout{** WARNING: IEEEtran.bst: No hyphenation pattern has been}%
\typeout{** loaded for the language `#1'. Using the pattern for}%
\typeout{** the default language instead.}%
\else
\language=\csname l@#1\endcsname
\fi
#2}}

\bibitem{Sha:61:BSPS}
C.~E. Shannon, ``Two-way communication channels,'' in \emph{Proc.\ 4th Berkeley
  Symp.\ Probability and Statistics}, vol.~1, Berkeley, CA, 1961, pp. 611--644.

\bibitem{RW:07:JSAC}
B.~Rankov and A.~Wittneben, ``Spectral efficient protocols for half-duplex
  fading relay channels,'' \emph{{IEEE} J. Select. Areas Commun.}, vol.~25,
  no.~2, pp. 379--389, Feb. 2007.

\bibitem{PY:07:ICC}
P.~Popovski and H.~Yomo, ``Physical network coding in two-way wireless relay
  channels,'' in \emph{Proc.\ IEEE Int.\ Conf.\ Communications (ICC'07)},
  Glasgow, Scotland, June 2007, pp. 707--712.

\bibitem{KMT:08:IT}
S.~J. Kim, P.~Mitran, and V.~Tarokh, ``Performance bounds for bidirectional
  coded cooperation protocols,'' \emph{{IEEE} Trans. Inform. Theory}, vol.~54,
  no.~11, pp. 5235--5241, Nov. 2008.

\bibitem{OBSB:08:IT}
T.~J. Oechtering, I.~Bjelakovic, C.~Schnurr, and H.~Boche, ``Broadcast capacity
  region of two-phase bidirectional relaying,'' \emph{{IEEE} Trans. Inform.
  Theory}, vol.~54, no.~1, pp. 454--458, Jan. 2008.

\bibitem{SSO:08:ITW}
C.~Schnurr, S.~Stanczak, and T.~J. Oechtering, ``Achievable rates for the
  restricted half-duplex two-way relay channel under a
  partial-decode-and-forward protocol,'' in \emph{Proc.\ IEEE Information
  Theory Workshop (ITW'08)}, Porto, Portugal, May 2008, pp. 134--138.

\bibitem{CHK:08:ICC}
T.~Cui, T.~Ho, and J.~Kliewer, ``Memoryless relay strategies for two-way relay
  channels: Performance analysis and optimization,'' in \emph{Proc.\ IEEE Int.\
  Conf.\ Communications (ICC'08)}, Beijing, China, May 2008, pp. 1139--1143.

\bibitem{Gal:68:Book}
R.~G. Gallager, \emph{Information Theory and Reliable Communication}.\hskip 1em
  plus 0.5em minus 0.4em\relax New York: Wiley, 1968.

\bibitem{Ahm:97:PhD}
W.~K.~M. Ahmed, ``Information theoretic reliability function for flat fading
  channels,'' Ph.D. dissertation, Queen's University, Kingston, ON, Canada,
  Sept. 1997.

\bibitem{AM:99:IT}
W.~K.~M. Ahmed and P.~J. McLane, ``Random coding error exponents for
  two-dimensional flat fading channels with complete channel state
  information,'' \emph{{IEEE} Trans. Inform. Theory}, vol.~45, no.~4, pp.
  1338--1346, May 1999.

\bibitem{SW:05:COM}
\BIBentryALTinterwordspacing
H.~Shin and M.~Z. Win, ``Gallager's exponent for {MIMO} channels: A
  reriability--rate tradeoff,'' \emph{{IEEE} Trans. Commun.}, to be published.
  [Online]. Available: \url{http://arxiv.org/abs/cs/0607095}
\BIBentrySTDinterwordspacing

\bibitem{HA:03:WCOM}
M.~O. Hasna and M.-S. Alouini, ``End-to-end performance of transmission systems
  with relays over {R}ayleigh-fading channels,'' \emph{{IEEE} Trans. Wireless
  Commun.}, vol.~2, no.~6, pp. 1126--1131, Nov. 2003.

\bibitem{HA:04:COM}
------, ``Harmonic mean and end-to-end performance of transmission systems with
  relays,'' \emph{{IEEE} Trans. Commun.}, vol.~52, no.~1, pp. 130--135, Jan.
  2004.

\bibitem{AK:04:WCOM}
P.~A. Anghel and M.~Kaveh, ``Exact symbol error probability of a cooperative
  network in a {R}ayleigh-fading environment,'' \emph{{IEEE} Trans. Wireless
  Commun.}, vol.~3, no.~5, pp. 1416--1421, Sept. 2004.

\bibitem{MS:78:Book}
A.~M. Mathai and R.~K. Saxena, \emph{The $H$-function with Applications in
  Statistics and Other Disciplines}.\hskip 1em plus 0.5em minus 0.4em\relax New
  York: Wiley, 1978.

\bibitem{BV:04:Book}
S.~Boyd and L.~Vandenberghe, \emph{Convex Optimization}.\hskip 1em plus 0.5em
  minus 0.4em\relax Cambridge, UK: Cambridge University Press, 2004.

\bibitem{LVBL:98:LAA}
M.~S. Lobo, L.~Vandenberghe, S.~Boyd, and H.~Lebret, ``Applications of
  second-order cone programming,'' in \emph{Linear Algebra and Its Appl.}, vol.
  284, Nov. 1998, pp. 193--228.

\bibitem{GR:00:Book}
I.~S. Gradshteyn and I.~M. Ryzhik, \emph{Table of Integrals, Series, and
  Products}, 6th~ed.\hskip 1em plus 0.5em minus 0.4em\relax San Diego, CA:
  Academic, 2000.

\bibitem{PBM:90:Book:v3}
A.~P. Prudnikov, Y.~A. Brychkov, and O.~I. Marichev, \emph{Integrals and
  Series}.\hskip 1em plus 0.5em minus 0.4em\relax New York: Gordon and Breach
  Science, 1990, vol.~3.

\end{thebibliography}

\clearpage

\clearpage

\clearpage

\clearpage

\clearpage

\clearpage

\clearpage

\clearpage

\clearpage

\end{document}